\documentclass[portrait,iop,revtex4]{emulateapj}

\newcommand{\gta}{\lower 0.5ex\hbox{$ \buildrel>\over\sim\ $}}
\newcommand{\lta}{\lower 0.5ex\hbox{$ \buildrel<\over\sim\ $}}
\newcommand{\nhe} {$N$({\rm He})/$N$({\rm H})}
\newcommand{\teff}{$T_{\rm eff}$}

\usepackage{natbib}
\usepackage{graphicx}
\usepackage{url}
\usepackage{amsmath}

\begin{document}

\title{Improved determination of the atmospheric parameters of the
  pulsating sdB star Feige 48} 

\author{M. Latour\altaffilmark{1}, G. Fontaine\altaffilmark{1},
   E.M. Green\altaffilmark{2}, P. Brassard\altaffilmark{1}, and
  P. Chayer\altaffilmark{3}}

\altaffiltext{1}{D\'epartement de Physique, Universit\'e
  de Montr\'eal, Succ. Centre-Ville, C.P. 6128, Montr\'eal, QC H3C 3J7,
  Canada} 
\altaffiltext{2}{Steward Observatory, University of Arizona, 933 North
  Cherry Avenue, Tucson, AZ 85721}  
\altaffiltext{3}{Space Telescope Science Institute, 3700 San Martin
  Drive, Baltimore, MD 21218}

\begin{abstract}
As part of a multifaceted effort to exploit better the
asteroseismological potential of the pulsating sdB star Feige 48, we
present an improved spectroscopic analysis of that star based on new
grids of NLTE, fully line-blanketed model atmospheres. To that end, we
gathered four high S/N time-averaged optical spectra of varying spectral
resolution from 1.0 \AA~to 8.7 \AA, and we made use of the results of four
independent studies to fix the abundances of the most
important metals in the atmosphere of Feige 48.
The mean atmospheric parameters we obtained from our
four spectra of Feige 48 are : \teff~= 29,850 $\pm$ 60 K, log $g$ = 5.46
$\pm$ 0.01, and log \nhe~= $-$2.88 $\pm$ 0.02.  We also modeled for the
first time the He~\textsc{ii} line at 1640 \AA~from the STIS archive
spectrum of the star and we found with this line an effective 
temperature and a surface gravity that match well the values obtained
with the optical data. With some fine tuning of the abundances of the
metals visible in the optical domain we were able to achieve a 
very good agreement between our best available spectrum 
and our best-fitting synthetic one. Our derived
atmospheric parameters for Feige 48 are in rather good agreement with
previous estimates based on less sophisticated models. This
underlines the relatively small effects of the NLTE approach combined
with line blanketing in the atmosphere of this particular star,
implying that the current estimates of the atmospheric parameters of
Feige 48 are reliable and secure. 

\end{abstract}


\section{INTRODUCTION} 
\citet{koen98} first reported the discovery of short-period (340-380 s)
pulsations in the hot B subdwarf (sdB) star Feige 48. Since then, that
star has attracted attention because it is relatively bright ($V$ =
13.48) for a pulsator of the kind, and because its optical light curve
shows relatively large pulsation amplitudes that may reach a few percent
of the mean intensity of the star. This made it an ideal candidate for
follow-up studies aimed at ultimately exploiting its full asteroseismic
potential. 

As a pulsator, Feige 48 was identified by \citet{koen98} to the then
newly-found short-period $p$-mode oscillators of the EC 14026 type
discovered shortly before \citep{kilkenny1997, koen1997, stobie1997, 
o1997}. Interestingly, the existence of this class of pulsators had been
predicted independently by theory \citep{char1996, char1997}. These
pulsators are now officially known as V361 Hya stars, and informally
referred to as sdB$_r$ stars \citep{kil2010}. There exists another
category of pulsating sdB's, the long-period $g$-mode pulsators of the
V1093 type (or sdB$_s$) discovered by \citet{green2003}. \citet{fon2003}
showed that the same basic process, a $\kappa$-mechanism fed by
radiative levitation of iron-peak elements, is responsible for the
excitation of pulsation modes in both types of pulsating hot B
subdwarfs. Feige 48 finds itself at the common boundary between the
hotter, higher gravity sdB$_r$ stars and the cooler, less compact
sdB$_s$ pulsators in the log $g$-$T_{\rm eff}$ diagram (see, e.g.,
Fig. 1 of \citealt{charp2013}). 

These two families of pulsating hot subdwarf stars show strong
similarities with the pair $\beta$-Cephei / Slowly Pulsating Blue stars
on the main sequence. However, sdB's are evolved stars that lie well
below the main sequence in the Herzsprung-Russell diagram on the
so-called Extreme Horizontal Branch. They are hot (22,000 K $\leq$
$T_{\rm eff}$ $\leq$ 38,000 K), compact (5.2 $\leq$ log $g$ $\leq$ 6.2)
core helium burning objects. The hot B subdwarf stars are also chemically
peculiar, showing strong He deficiencies and unusual metal abundance
patterns (see, e.g., \citealt{geier13}). \citet{heber2009} provides a
comprehensive review of the properties of these intrinsically
interesting, but still often neglected stars.

On the asteroseismological front, follow-up photometric observations
gathered by \citet{reed2004} with small telescopes over a period of five
years confirmed the initial detection of five pulsation modes in Feige
48 as reported by \citet{koen98}. Independent observations 
obtained at the Canada-France-Hawaii Telescope (CFHT) using the
Montr\'eal 3-channel photometer LAPOUNE
produced a significant improvement of
sensitivity leading to the uncovering of nine distinct pulsation modes,
including the five previously known \citep{char05f48}. These modes were
found to belong to four multiplet structures (two triplets, one doublet,
and one singlet) associated with rotational splitting. For comparison
with nonrotating models, only the four central periods of these
complexes could be used, however. On this basis, \citet{char05f48}
presented a  preliminary seismic model of Feige 48 following the forward
method developed by \citet{brassard2001}. By incorporating rotation at
the outset in their models, \citet{vang08} were able to make full use of
the nine distinct pulsation modes previously detected in the CFHT/LAPOUNE
campaign and infer part of the internal rotation profile of Feige
48. This was used to test in a preliminary way spin-orbit synchronism in
the close binary system that Feige 48 belongs to. Indeed, using HST/STIS
observations and archive FUSE data, \citet{otoole04} had previously 
found that Feige 48 is a member of such a system, with an orbital period
of about 9 h. According to these authors, the unseen companion is likely
a white dwarf, although the hypothesis of a cool main sequence star
could not be completely ruled out. 

Given the intrinsic importance of testing spin-orbit synchronism in
close binary systems in general, we decided to exploit further the
opportunity offered by the Feige 48 system. To this end, we first
invested in a major white light photometric campaign from the ground,
with the main objective of detecting many more pulsation modes than
the nine uncovered previously. We were able to gather nearly 400 h of
very high S/N data using the Mont4K camera attached to the Kuiper
Telescope of the Steward Observatory Mount Bigelow Station. This was
highly successful as some 46 pulsation modes were detected in Feige
48. Details will be reported elsewhere by Green et al. (in
preparation). 

As an integral part of these efforts, which will culminate with a new
detailed seismic analysis of Feige 48 (Van Grootel et al., in
preparation), we have pursued a spectroscopic campaign to obtain
accurate radial velocity measurements and high S/N time-averaged
spectra. The latter were used to obtain a more accurate determination
of the atmospheric parameters of Feige 48, and this is what we report in
this paper. As shown specifically in both \citet{char05f48} and
\citet{vang08}, independent estimates of the atmospheric parameters of
the pulsator -- as provided by spectroscopy coupled to model atmosphere
calculations -- have been necessary to lift degeneracies in seismic
solutions. This operation was then deemed crucial to our present
multipronged efforts to understand better Feige 48 as a pulsator. We
briefly review below what has been done in the past in terms of
atmospheric analyses of the sdB star Feige 48, and we present the
results of our own efforts based on the combination of the NLTE approach
with the inclusion of detailed metal line blanketing in the atmosphere
models.

\section{SOME BACKGROUND}

Hot subdwarf stars span a wide range of effective temperatures,
from around 22,000 K for the coolest sdB's to almost 100,000 K in the
hottest sdO's. Depending on the effective temperature, model atmospheres 
of varying sophistication need to be used in order to determine in a
reliable way the atmospheric parameters and chemical composition of a hot 
subdwarf. For sdO stars, it has been shown that model atmospheres using
the local thermodynamic equilibrium (LTE) approximation fail to
reproduce observed spectra and lead to incorrect atmospheric parameters when
one tries to fit the observed Balmer and helium lines with this type of
model. In these stars, the non-LTE (NLTE) effects are undeniably
important and must be taken into account when modeling their atmospheric
layers. On the other hand, for the coolest sdB stars, the LTE approximation
usually gives correct results in terms of atmospheric parameters. The
NLTE effects begin to be nonnegligible in the hottest sdB's with
effective temperatures around and beyond 30,000 K, which is also the
temperature where He~\textsc{ii} lines become visible in the optical spectra
\citep{nap97}. For these hot sdB's (sometimes referred to as sdOB
stars), it becomes difficult to reproduce simultaneously both the
He~\textsc{i} and \textsc{ii} lines, as is the case, for example, of the
star PG 1219$+$534 where both LTE and NLTE approaches with metal-free
models fail to reproduce helium lines from both ionization stages
(\citealt{heb00} and \citealt{char05pg}). 

\begin{figure}[t]
\epsscale{1.2}
\plotone{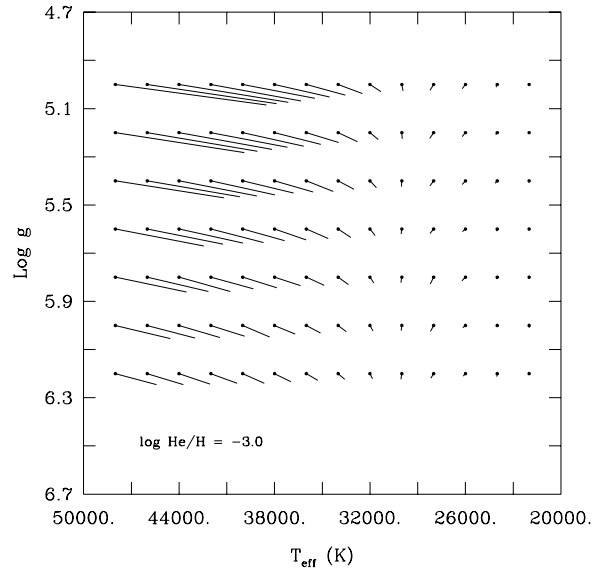}
\caption{NLTE effects in the log $g$-$T_{\rm
  eff}$ plane. The dots give the original values of the atmospheric
  parameters for NLTE models with log {\it N}(He)/{\it N}(H) = $-$3.0
  and no metals. The opposite end of each vector indicates the fictitious
  values of these parameters when the original spectra (treated as 
  observational data) are analyzed with a grid of LTE models (again with
  no metals).}
\label{mapnlte}
\end{figure}

The situation concerning the importance of NLTE effects in sdB stars is,
in fact, more complicated than the simple rule of thumb proposed by
\citet{nap97}, namely, that these effects are negligible below $T_{\rm
  eff}$ $\sim$ 30,000 K. There is also a dependency on the surface
gravity and on the helium content. Figure 1 illustrates the results of
analyzing a grid of synthetic spectra computed from metal-free NLTE
models and treated as ``observational'' data, with a grid of
theoretical spectra based on a similar grid of metal-free models, but
computed, this time, in the LTE approximation. The helium abundance in
this diagram refers to a value close to that found in Feige 48 (see
below). The dots indicate the original values of the parameters of the
NLTE models, while the end of each line segment indicates the fake values
of these parameters as inferred with the LTE models. As expected, the
largest deviations occur for the hotter and less compact atmospheres. 
And while it is true that the magnitude of the NLTE vectors becomes
relatively small below $T_{\rm eff}$ $\sim$ 30,000 K, the behavior is
complex and not simply monotonic as revealed by the ``rotating''
vectors.

\clearpage
\begin{figure}[t]
\epsscale{1.2}
\plotone{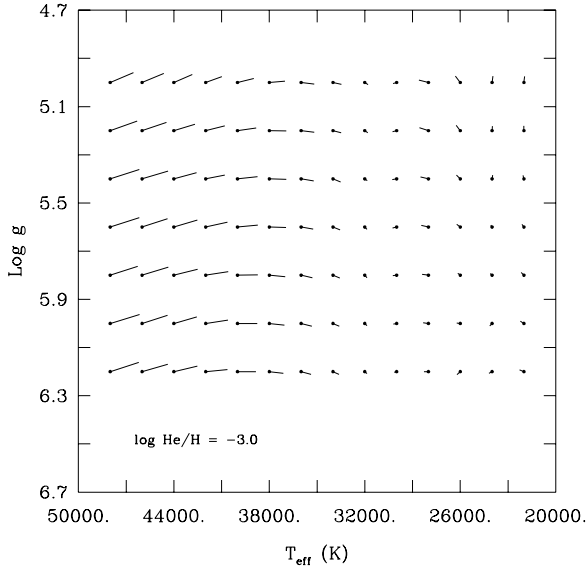}
\caption{Metal blanketing effects in the log
  $g$-$T_{\rm eff}$ plane. The dots give the original values of the
  atmospheric parameters for NLTE models with log {\it N}(He)/{\it N}(H)
  = $-$3.0 and a metallicity specified by C(0.1 solar), N(solar), O(0.1
  solar), Si (0.1 solar), S(solar), and Fe(solar). The opposite end of
  each vector indicates the fictitious values of these parameters when
  the original spectra (treated as observational data) are analyzed with
  a grid of NLTE models with no metals.}
\label{mapline}
\end{figure}

Another important ingredient in atmosphere modeling are the metallic
elements that cause line blanketing and thus change the shape of the
spectral energy distribution of the star. Changes in the emergent
spectrum are more important in the UV domain and at shorter wavelengths
where a large number of absorption lines, mainly from iron-peak elements, 
block an important fraction of the flux. The flux must nevertheless
come out of the star, and it does so at larger wavelengths, increasing
the continuum level of these regions. The line blanketing caused by
metals also has an effect on the thermodynamical structure of the
atmosphere by heating the inner layers and cooling the outer ones. 
All of this has repercussions in the optical domain where the
Balmer and helium lines used to determine the atmospheric parameters of
a star are found. If these lines are affected by the presence of metallic
elements, so will the atmospheric parameters derived by fitting
them. 

Adding elements heavier than helium into model atmospheres computed in
the LTE approximation is now a common practice and LTE line-blanketed
model atmospheres are often used to study sdB stars. As for NLTE models,
including metals in them is a much more complicated task than doing it
in their LTE counterparts. As a consequence, and to our knowledge, the
only extensive and detailed grids of atmosphere models for sdB stars
combining the NLTE approach with metals have been those of \citet{bra10} and
\citet{nemeth2012}. Figure 2 based on some of the models of
\citet{bra10} illustrates particularly well the effects of including C,
N, O, Si, S, and Fe in NLTE calculations for sdB stars on the derived
atmospheric parameters.
Those elements are the most abundant metals that
have been measured in a sample of five sdB$_s$ stars by \citet{blan2008}
on the basis of UV data from the Far-Ultraviolet Spectroscopic Explorer (FUSE),
and their average abundances have been
adopted as a representative metallicity in the NLTE calculations of
\citet{bra10}. Given that previous estimates of the atmospheric
parameters of Feige 48 have led to $T_{\rm eff}$ $\simeq$ 30,000 K and
log $g$ $\simeq$ 5.5 (see below), one can anticipate from Figure 2 that
the effects of metal line blanketing cannot be very large in the
atmosphere of that star. Still, given the importance of Feige 48 as a
pulsator in a close binary system, we have felt it worthwhile to carry
out the present analysis using the state of the art models in the field.

\section{SPECTRAL ANALYSIS OF FEIGE 48}

\subsection{Previous Studies}

The first estimates of the atmospheric parameters of Feige 48 based on a
spectroscopic analysis were obtained by \citet{koen98} when they found the
star to be an EC 14026 pulsator. They used a grid of LTE model
atmospheres with pure hydrogen, and this led to an effective temperature
of 28,900 $\pm$ 300 K and a log $g$ of 5.45 $\pm$ 0.05. Later,
\citet{heb00} revised these values a bit upward with their detailed
analysis of a Keck/HIRES spectrum of Feige 48. They derived atmospheric
parameters by fitting the Balmer and He~\textsc{i} lines with three kinds of
models: LTE with a solar metalicity, metal-poor LTE with [M/H] = $-$2.0,
and NLTE without metals (H and He only). Their temperature estimates
agreed well with each other, and they were slightly higher than the one
found by \citet{koen98}. Taking the mean value of their three estimates
and the one from \citet{koen98}, they finally adopted 29,500 $\pm$ 350
K as the effective temperature of Feige 48. Their values of log $g$
obtained with the three types of models are similar, and using the 
same approach as they did for the temperature, \citet{heb00} got a mean
log $g$ of 5.50 $\pm$ 0.05. Finally, the helium abundance was basically
also the same for each type of models, and they got a final value of log
\nhe~= $-$2.93 $\pm$ 0.05. Note that the Keck/HIRES data allowed Heber
et al. (2000) to constrain the projected rotational velocity of Feige 48 
to v$_{rot}$sin i $\leq$5 km s$^{-1}$, thus indicating that the star is a 
relatively slow rotator or an object
seen nearly pole-on, or both.

When Feige 48 was studied from an asteroseismological point of view by
\citet{char05f48}, these authors also obtained a new, independent
derivation of its atmospheric parameters. \citet{char05f48} combined a
high S/N, medium resolution ($\sim$1 \AA) spectrum that they gathered at
the MMT with a grid of metal-free H,He NLTE model atmospheres, obtaining
\teff~= 29,580 $\pm$ 370 K, log $g$ = 5.48 $\pm$ 0.05, and 
log \nhe~= $-$2.95 $\pm$ 0.08. Their asteroseismic analysis confirmed the
spectroscopic value of the gravity with a best match at log $g$ = 5.44.  
Hence, according to the previous available spectroscopic analyses, the
atmospheric parameters of Feige 48 seem to be well constrained and do
not show an important dependence on either the type of model atmospheres
(LTE or NLTE) or the presence of metallic elements in LTE models. 
 
In the past few years, we developed an efficient way of computing NTLE
line-blanketed model atmospheres using versions of the
public codes TLUSTY and SYNSPEC (\citealt{lanz95,lanz03,lanz07}) running on a
cluster of dedicated PC's that now include 320 processors (see
\citealt{lat11} for more details on the setup). 
More than 300 model atmospheres can thus be computed at the same
time.
This made it
straightforward for us to use this more sophisticated approach
for modeling the atmosphere of Feige 48. This allows for exploiting 
at their full potential the spectroscopic data that we gathered on 
Feige 48, and for supporting its forthcoming seismic analysis with the 
most accurate atmospheric parameters that can be currently obtained.

\subsection{Observational Material}

We have at our disposal four optical spectra of Feige 48, spanning
different wavelength ranges and having different resolutions. Our best
data set is a combination of medium resolution (1 \AA) spectra acquired
between 2002 and 2013 with the blue spectrograph attached to the 6.5 m
Multiple Mirror Telescope (MMT). This was part of the ongoing radial
velocity program on sdB stars carried out by one of us (E.M.G.). 
Throughout the observing seasons, the same experimental setup was
consistently used. The 832 mm$^{-1}$ grating in second order and 1'' slit
provide a resolution $R$ of $\sim$4250 (1.05 \AA) over the wavelength
range 4000--4950 \AA. The slit was always aligned at the parallactic
angle during the observations. Exposures of 240 to 475 s, depending on
conditions, resulted in signal-to-noise ratios (S/N) of about 80 to
150 for individual spectra, which is sufficient to achieve velocity
errors of 1 to 2 km s$^{-1}$ for sdB stars. One to four spectra per night were
obtained. The spectrum used for the present spectroscopic analysis is
the combination of 17 mean nightly spectra gathered during different
runs, median-filtered and shifted to the same velocity prior to combining.
Thus the resulting spectrum has a remarkably high formal S/N of $\sim$460. This
spectrum will be referred to as MMT. We point out that an earlier,
reduced sensitivity version of this spectrum was used by
\citet{char05f48} in their seismic analysis.  

Our second spectrum is also a medium-resolution one obtained
from the combination of several individual spectra gathered over the
last several years within the context of the radial velocity program of
E.M.G. carried out on sdB stars. These data were obtained with the
Boller \& Chivens (B\&C) Cassegrain spectrograph at Steward
Observatory's 2.3 m Bok Telescope on Kitt Peak. The 832 mm$^{-1}$
grating in second order with a 1.5'' slit were used to achieve 1.9
\AA~resolution over a wavelength range of 3675--4520 \AA. The slit was
aligned with the parallactic angle at the midpoint of each exposure, and
comparison HeAr spectra were taken before and after each stellar
spectrum. Exposure times between 500 and 875 s for the individual
spectra led to S/N of about 50 to 80. The final spectrum is the
combination of 50 exposures and has a resulting S/N of
$\sim$375. This spectrum will be referred to as BG2 (as in ``Betsy
Green's $\sim$2 \AA~spectrum'').

Another spectrum of Feige 48 was kindly gathered for us by Pierre
Bergeron using again the B\&C spectrograph on the 2.3 m Bok
telescope. This was part of a request to observe for us several
pulsating sdB stars during two of his white dwarf observing runs going
back to 2006. In that case, the 4.5'' slit together with the 600 mm$^{-1}$
grating blazed at 3568 \AA~ in first order provided a spectral coverage
from about 3030 to 5250 \AA~ at a resolution of $\sim$6.0 \AA. The S/N
of this single spectrum is $\sim$80. Even though this spectrum is of
lower quality than the previous ones in terms of sensitivity and
resolution, we felt that it would be worthwhile to analyze all spectral
data available to us on our target star. This third spectrum is referred
to as PB6 in what follows.

Finally, we have at our disposal an older (2004) high-sensitivity,
low-resolution spectrum of Feige 48 gathered again with the B\&C
spectrograph on the 2.3 m Bok telescope. It is the combination of 5
exposures using the 400 mm$^{-1}$ grating in first order in conjunction
with a 2.5'' slit to obtain a typical resolution of 8.7 \AA~over the
wavelength interval 3620--6900 \AA, thus including H$\alpha$. The
instrument rotator was set prior to each exposure, to align the slit
within $\sim$2$^{\circ}$ of the parallactic angle at the midpoint of the
exposure. HeAr comparison spectra were obtained immediately following
each stellar exposure. The blue part of the combined
spectrum reaches S/N $\simeq$ 248. This spectrum is to be referred to
as BG9. It was used previously by \citet{char05f48} in their analysis of
Feige 48.    

Inspection of our optical data revealed no sign of He~\textsc{ii} lines
in the spectra. There is also no hint for the presence of
He~\textsc{ii} $\lambda$4686 or other weaker features associated with
that ionization stage in the high-resolution HIRES spectrum of
Feige 48 obtained by \citet{heb00}. However, since Feige 48 has been
observed with the STIS spectrograph and these data are available in
the Mikulski Archive for Space Telescopes
(MAST)\footnote{http://archive.stsci.edu/}, we checked the UV spectrum
in order to verify if the strongest expected He~\textsc{ii} feature in
that wavelength range, the $\lambda$1640 line, could be detected (see
\citealt{otoole06} for more details on the data). Indeed, the line is
present, although somewhat weak and noisy, but it is still sufficiently
useful as the sole indicator of that ionization stage of helium to
provide us below with a nice test of the validity of our derived
atmospheric parameters based on the optical data. 

\subsection{Model Atmospheres}

\subsubsection{The Metallicity of Feige 48}

\begin{deluxetable*}{cccccccc}[t!]
\tablewidth{0pt}
 \tabletypesize{\scriptsize}
\tablecaption{Abundances of Metals Detected in the Atmosphere of
  Feige 48 : $\log$ $N$(Z)/$N$(H) } 
\tablehead{
\colhead{Element} &
 \colhead{\citet{heb00}} &
 \colhead{\citet{chay04}} &
 \colhead{\citet{otoole06}} &
 \colhead{\citet{geier13}} &
\colhead{Mean} &
 \colhead{This work} &
\colhead{Solar\tablenotemark{a}} \\
\colhead{Z} &
 \colhead{Keck/HIRES} &
 \colhead{$FUSE$} & 
 \colhead{$HST$/STIS} &
 \colhead{Keck/HIRES} &
\colhead{} &
 \colhead{MMT} &
\colhead{}
}
\startdata
\textbf{C} & $-$4.64$\pm$0.03 & $-$5.2$\pm$0.5 & $-$4.79$\pm$0.10 & $-$4.65$\pm$0.35
& \textbf{$-$4.65$\pm$0.03} & $-$4.94$\pm$0.1 &$-$3.57 \\
\textbf{N} & $-$4.29$\pm$0.10 & $-$4.6$\pm$0.5 & $-$4.38$\pm$0.38 & $-$4.72$\pm$0.12
& \textbf{$-$4.47$\pm$0.07} & $-$4.61$\pm$0.2 &$-$4.17 \\
\textbf{O} & $-$4.21$\pm$0.12 & $<-$4.2 & ... & $-$4.35$\pm$0.20
& \textbf{$-$4.25$\pm$0.10} & $-$4.47$\pm$0.1 &$-$3.31 \\
\textbf{Ne} & $-$4.90$\pm$0.31 & ... & ... & $<-$4.06
& \textbf{$-$4.90$\pm$0.31} & ... &$-$4.07 \\
\textbf{Mg} & $-$5.09$\pm$0.50 & ... & ... & $-$5.2$\pm$0.50
& \textbf{$-$5.15$\pm$0.35} & $-$5.18$\pm$0.2 & $-$4.40 \\
Al & $-$5.50$\pm$0.18 & $<-5.8$ & $-$6.49$\pm$0.10 & $-$6.4$\pm$0.5
& $-$6.49$\pm$0.09 & ... & $-$5.55 \\
\textbf{Si} & $-$5.67$\pm$0.27 & $-$5.7$\pm$0.5 & $-$5.73$\pm$0.19 & $-$5.45$\pm$0.06
& \textbf{$-$5.49$\pm$0.06} & $-$5.38$\pm$0.3 & $-$4.49 \\
P & ... & $-$7.1$\pm$0.5 & $-$7.59$\pm$0.50 & $<-7.02$ & $-$7.35$\pm$0.35 
& ... & $-$6.59 \\
S & $-$5.85$\pm$0.50 & $-$5.7$\pm$0.5 & ... &  $-$6.05$\pm$0.07
& $-$6.04$\pm$0.07 & $-$5.71$\pm$0.3 & $-$4.88 \\
Cl & ... & $<-8.5$ & ... &  ... & $<-8.5$ & ... &$-$6.50 \\
Ar & ... & $<-6.8$ & $<-5.0$ & $<-5.18$ & $<-6.8$ & ... & $-$5.60 \\
K & ... & ... & ... & $<-6.07$ & $<-6.07$ & ... &$-$6.97 \\
Ca & ... & ... & $-$4.69$\pm$0.50 & ... & $-$4.69$\pm$0.50 & ... & $-$5.66 \\
Sc & ... & ... & $<-9.0$ & ... & $<-9.0$ & ... & $-$8.85 \\
Ti & ... & $<-7.6$ & $-$6.81$\pm$0.13 & $<-6.26$ & $-$6.81$\pm$0.13
& ... &$-$7.05 \\
V & ... & $<-8.3$ & ... & $<-5.03$ & $<-8.3$ & ... &$-$8.07 \\
Cr & ... & $-$7.0$\pm$0.5 & $-$5.95$\pm$0.15 & ... & $-$6.04$\pm$0.14
& ... &$-$6.36 \\
Mn & ... & $-$7.0$\pm$0.5 & $-$6.38$\pm$0.21 & ... & $-$6.47$\pm$0.19
& ... &$-$6.57 \\
\textbf{Fe} & $-$4.45$\pm$0.19 & $-$4.8$\pm$0.5 & $-$4.30$\pm$0.14 & $-$4.54$\pm$0.21
& \textbf{$-$4.41$\pm$0.10} & $-$4.48$\pm$0.2 &$-$4.50 \\
Co & ... & $-$7.6$\pm$0.5 & $-$6.11$\pm$0.18 & ... & $-$6.28$\pm$0.17
& ... &$-$7.01 \\
\textbf{Ni} & ... & $<-$5.7 & $-$5.31$\pm$0.15 & ...
& \textbf{$-$5.31$\pm$0.15} & ... &$-$5.78 \\
Cu & ... & ... & $-$6.75$\pm$0.40 & ... & $-$6.75$\pm$0.40 & ... &$-$7.81 \\
Zn & ... & ... & $-$6.70$\pm$0.22 & ... & $-$6.70$\pm$0.22 &  ... &$-$7.44 \\
Ga & ... & ... & $-$7.30$\pm$0.50 & ... & $-$7.30$\pm$0.50 & ... & $-$8.96 \\
Ge & ... & ... & $-$7.98$\pm$0.08 & ... & $-$7.98$\pm$0.08 & ... & $-$8.35 \\
Sn & ... & ... & $-$9.06$\pm$0.50 & ... & $-$9.06$\pm$0.50 & ... & $-$9.96 \\
Pb & ... & ... & $-$8.20$\pm$0.50 & ... & $-$8.20$\pm$0.50 & ... & $-$10.25 \\
\enddata
\tablenotetext{a}{Asplund et al. (2009)}
\end{deluxetable*}

In order to fix a suitable chemical composition for the grid of models, we
searched the literature for abundance studies done on Feige 48. Because
of its status of pulsating star and its brightness, Feige 48
has been thoroughly studied and its atmospheric chemical composition has been
analyzed in at least four different studies that we know of. First,
after their determination of the atmospheric parameters of the star on
the basis of a HIRES spectrum, \citet{heb00} carried out an abundance
analysis of visible metals and found the star to have subsolar
abundances for all of the eight elements they studied, except for iron
which was found to be solar. For their part, \citet{chay04} made a
comparison of the chemical composition of Feige 48 and Feige 87 (a
non-variable sdB with atmospheric parameters very similar to those of
Feige 48) using UV metallic lines present in the star's FUSE
spectrum. \citet{otoole06} also undertook an abundance analysis of a
sample of pulsating and constant sdB stars (among them Feige 48) in
order to test the hypothesis that pulsating sdB's might show a different
abundance pattern than the nonpulsating ones. This was not the case, but
they nevertheless got a rather good picture of the abundance patterns in
the five stars they studied. Finally, in an attempt to go further into
deriving general trends for the metallic abundances in sdB stars,
\citet{geier13} analyzed a much larger sample of sdB's (among them Feige
48) using high-resolution optical spectra. Among other results, he
obtained a fourth set of metal abundances for the star of interest. 

Table 1 reports our compilation of the abundances obtained in the four
previously mentioned studies. A weighted mean abundance was computed
for each element, and since the uncertainty is used as weight, we
attributed a value of $\pm$0.5 dex whenever there was no uncertainty
quoted in the reference paper. We included in our model grid the mean abundance of 
the eight metals that are indicated in bold in Table 1. 
We limited ourselves to
the inclusion of ten atomic species (including H and He) in our models,
for stability and convergence reasons. Therefore, we chose to include
the eight most abundant metallic species, except for calcium which we
left aside because its abundance is based on a single optical line.  
This table also features (7$^{th}$ column) the abundances that we derived
using the MMT spectra (see Section 3.5 below).

It is worth mentioning that \citet{heb00}, \citet{otoole06}, and
\citet{geier13} used the following parameters for their abundance
analyses: \teff~= 29,500 K, log $g$ = 5.54, and log \nhe~= $-$2.9. The
remaining study of \citet{chay04} used parameters quite similar. 
\citet{otoole06} used line-blanketed LTE model atmospheres with solar
metallicity. Otherwise, the three other studies also used LTE models
(although it is not clearly mentioned in \citealt{heb00}) with a certain
amount of metallic elements that was not explicitly mentioned.  

\subsubsection{Model Grids}

Having determined in Table 1 the metallicity to be included in our models, we
first built a small grid of 150 NLTE line-blanketed model atmospheres
especially suited for Feige 48. The grid includes 5 values of
\teff~between 26,000 K and 34,000 K in steps of 2000 K, 6 values of log
$g$ between 5.0 and 6.0 in steps of 0.2 dex, and 5 helium abundances
between log \nhe = $-$4.0 and $-$2.0 in steps of 0.5 dex. Our models
contains the following species: H, He, C, N, O, Ne, Mg, Si, Fe,
and Ni. The main characteristics of the model atoms used are
listed in Table 2. The population of each level and superlevel is computed
in NLTE and the number of lines correspond to the allowed
transitions between these levels and superlevels.
Model atoms of iron and nickel are built in a different way than the lighter elements: all the NLTE levels are in fact superlevels and each of them includes a certain number of individual levels. The populations of individual levels inside a superlevel are computed assuming a Boltzmann distribution. The number of line transitions indicated for iron and nickel are those occuring between the individual levels. Further technical details on the atomic data used in TLUSTY can be found in \citet{lanz03atom}.
The highest ionization stage of each element is taken as a one-level atom.
Most of the models we used are available on the TLUSTY web 
page\footnote{http://nova.astro.umd.edu/Tlusty2002/tlusty-frames-data.html}. 

\begin{deluxetable}{lcc}[t]
\tablewidth{0pt}
 \tabletypesize{\scriptsize}
\tablecaption{Details of the model atoms used in our model atmospheres}
\tablehead{
\colhead{Ion} &
\colhead{No. of levels/superlevels} &
\colhead{No. of lines\tablenotemark{a}} 
}
\startdata
\hline
H~\textsc{i}   & 8/1 & 30 \\
He~\textsc{i}   & 19/5 &  104 \\
He~\textsc{ii}   & 20/- & 121 \\
C~\textsc{ii}   & 17/5 &  105 \\
C~\textsc{iii}   & 34/12 & 380 \\
C~\textsc{iv}   & 21/4 & 150 \\
C~\textsc{v}   & 1 & - \\
N~\textsc{ii}   & 32/10 & 278 \\
N~\textsc{iii}   & 25/7 & 184 \\
N~\textsc{iv}   & 34/14 & 384 \\
N~\textsc{v}   & 10/6 & 86 \\
N~\textsc{vi}   & 1 & - \\
O~\textsc{ii}   & 36/12 &  346\\
O~\textsc{iii}   & 28/13 &  224 \\
O~\textsc{iv}   & 31/8 &  270 \\
O~\textsc{v}   & 34/6 &  222 \\
O~\textsc{vi}   & 1 & - \\
Ne~\textsc{i}  & 23/12 &  219  \\
Ne~\textsc{ii}   & 23/9 &  178 \\
Ne~\textsc{iii}   & 12/2 & 20 \\
Ne~\textsc{iv}   & 10/2 &  22 \\
Ne~\textsc{v}   & 1 & - \\
Mg~\textsc{i}  & 12/6 &  56 \\
Mg~\textsc{ii}   & 21/4 &  146 \\
Mg~\textsc{iii}   & 1 & - \\
Si~\textsc{ii}   & 13/3 &  48 \\
Si~\textsc{iii}   & 31/15 & 341 \\
Si~\textsc{iv}   & 19/4 & 130  \\
Si~\textsc{v}   & 1 & - \\
\hline
  & No. of levels NLTE/LTE & No. of lines\tablenotemark{b} \\
\hline
Fe~\textsc{ii}   & 36/10921 & 1264969 \\
Fe~\textsc{iii}   & 50/12660 & 1604934\\
Fe~\textsc{iv}   & 43/13705 & 1776984 \\
Fe~\textsc{v}   & 42/11989 & 1008385\\
Fe~\textsc{vi}   & 1 & - \\
Ni~\textsc{iii}   & 36/11335 & 1309729 \\
Ni~\textsc{iv}   & 38/13172 & 1918070 \\
Ni~\textsc{v}   & 48/13184 & 1971819 \\
Ni~\textsc{vi}   & 1 & - \\
\enddata
\tablenotetext{a}{Allowed transitions}
\tablenotetext{b}{Transitions between LTE levels}
\end{deluxetable}

\begin{figure}[t]
\epsscale{1.2}
\plotone{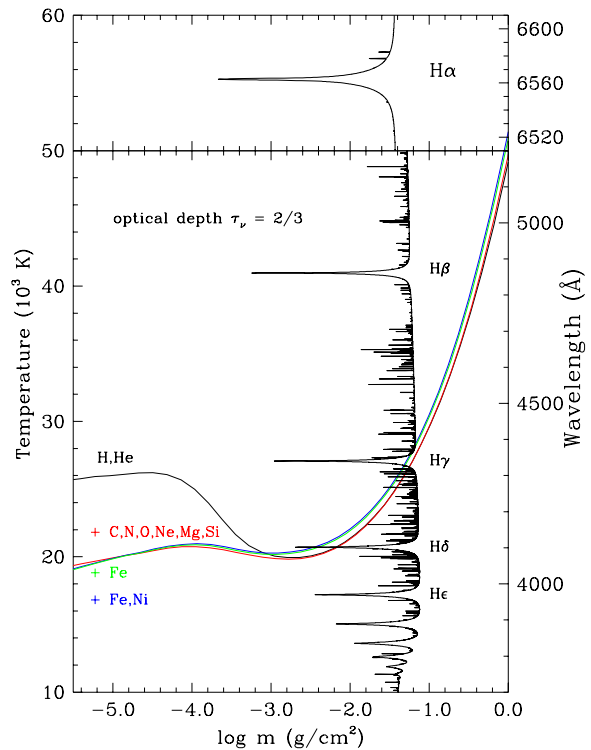}
\caption{Temperature stratification and monochromatic 
optical  depth $\tau_{\nu}$ = 2/3 as functions of depth, where {\it m}
is the column  density, for NLTE models defined by ${\it T}_{\rm eff}$ =
30,000 K, log $g$ = 5.4, and log {\it N}(He)/{\it N}(H) = $-$3.0. The
temperature structure is shown for four model atmospheres having
different compositions : with H and He only (black), with C, N, O, Ne,
Mg, Si in addition (red), with Fe added to the previously mentioned
elements (green), and finally with Ni on top of that (blue). The
$\tau_{\nu}$ = 2/3 curve is from the latter model and shows wavelength
intervals corresponding to the Balmer line series. }
\label{struc}
\end{figure}

Before performing the fitting procedure on
the spectra of Feige 48, we examined a few properties of our models.  
We first looked at the differences produced on the temperature structure
by the metallic elements included in the atmosphere as compared to the
metal-free case. Figure~\ref{struc} shows the temperature structures for
four model atmospheres having the same parameters (\teff = 30,000 K, log
$g$ = 5.4, and log \nhe = $-$3.0), except for their metallicity. The
black curve is from a model including only hydrogen and helium, showing
the typical NLTE temperature inversion in the upper layers of a
metal-free atmosphere. The red curve illustrates the case where only our
six ``lighter" metals were included in the calculations. Their main
effect is to cool down the upper layers. The most noteworthy effect of
adding iron (green curve) is the concomitant back-warming of the
deeper layers, which results in a higher temperature at a given depth.  
We also included the temperature profile (blue curve) of a fully
line-blanketed model which includes all of the eight metallic
elements. The only difference between the blue and the green curves is
the addition of nickel and it can be seen that, at least for a model
representing Feige 48, the influence of this element on the
thermodynamical structure of the model remains rather small. Overall, 
the line-blanketing effects are qualitatively the same as what is seen
in model atmospheres at higher temperature (for sdO stars), but to a
lesser extent. The drop in the surface temperature is around 7,000 K for
our Feige 48 models, while it can be around 40,000 K in 80,000 K sdO
models (see, e.g., Fig. 1 of \citealt{lat2013}). In the deeper layers (log
$m$ $\gta$ $-$3.0), the rise of the temperature due to the metals is
around four percent. Also featured in Figure~\ref{struc} is the optical
depth $\tau_{\nu} = 2/3$ as a function of the column density ($m$),
allowing one to infer where, in the atmosphere, the Balmer and metallic
lines as well as the continuum are formed. 

\begin{figure}[t]
\epsscale{1.2}
\plotone{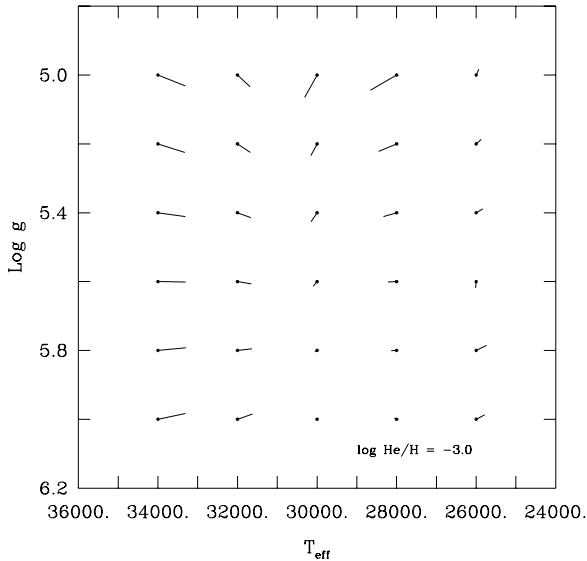}
\caption{Map illustrating the effects of metal line
blanketing on the inferred atmospheric parameters. \teff~and log $g$ are
obtained for some models of our NLTE line-blanketed grid fitted with a
grid of NLTE H,He models. The atmospheric parameters of the models are
indicated by dots while the end of line segments correspond to the
parameters obtained by the fitting procedure. The helium abundance is
kept fixed at the model value of log \nhe~=$-$3.0.}
\label{map}
\end{figure}

We also wanted to verify how these changes in the structure of our
models might affect the computed optical spectrum and, by extension, any
derived atmospheric parameters. To illustrate this, we created the map
featured in Figure~\ref{map}. This is similar to our Figure~\ref{mapline} above, but
specialized to the specific metallicity  used to model Feige 48. To
construct this map, spectra from one grid of models are considered as
``observed'' spectra and are fitted with a different grid of synthetic
spectra. We performed our fitting procedure in the 4020$-$4910~\AA\ interval, 
which includes three Balmer lines and four He~\textsc{i} lines, with
models convolved at 1~\AA\ resolution in order to mimic the MMT spectrum.
Because we wanted to check the effects of line blanketing, spectra from
our fully-blanketed grid were taken as the observed ones and we fitted
them with a grid of theoretical spectra obtained with model atmospheres
including only hydrogen and helium. Our $\chi^2$ fits were done in order
to derive an optimal solution in terms of effective temperature and surface
gravity, while the helium abundance was kept fixed at its real value. 
Figure~\ref{map} shows our results for a helium
abundance similar to the one found in Feige 48 (log \nhe~=$-$3.0). We
can see, once again and in complement to Figure 2, that the effects of
metals on the derived atmospheric parameters remain small throughout the
domain investigated. Somewhat accidentally in the specific case of Feige
48 with $T_{\rm eff}$ $\simeq$ 30,000 K and log $g$ $\simeq$ 5.5, the
effects are particularly small.

\subsection{Derived Atmospheric Parameters}

\begin{figure*}[t]
\includegraphics[scale=0.36,angle=270]{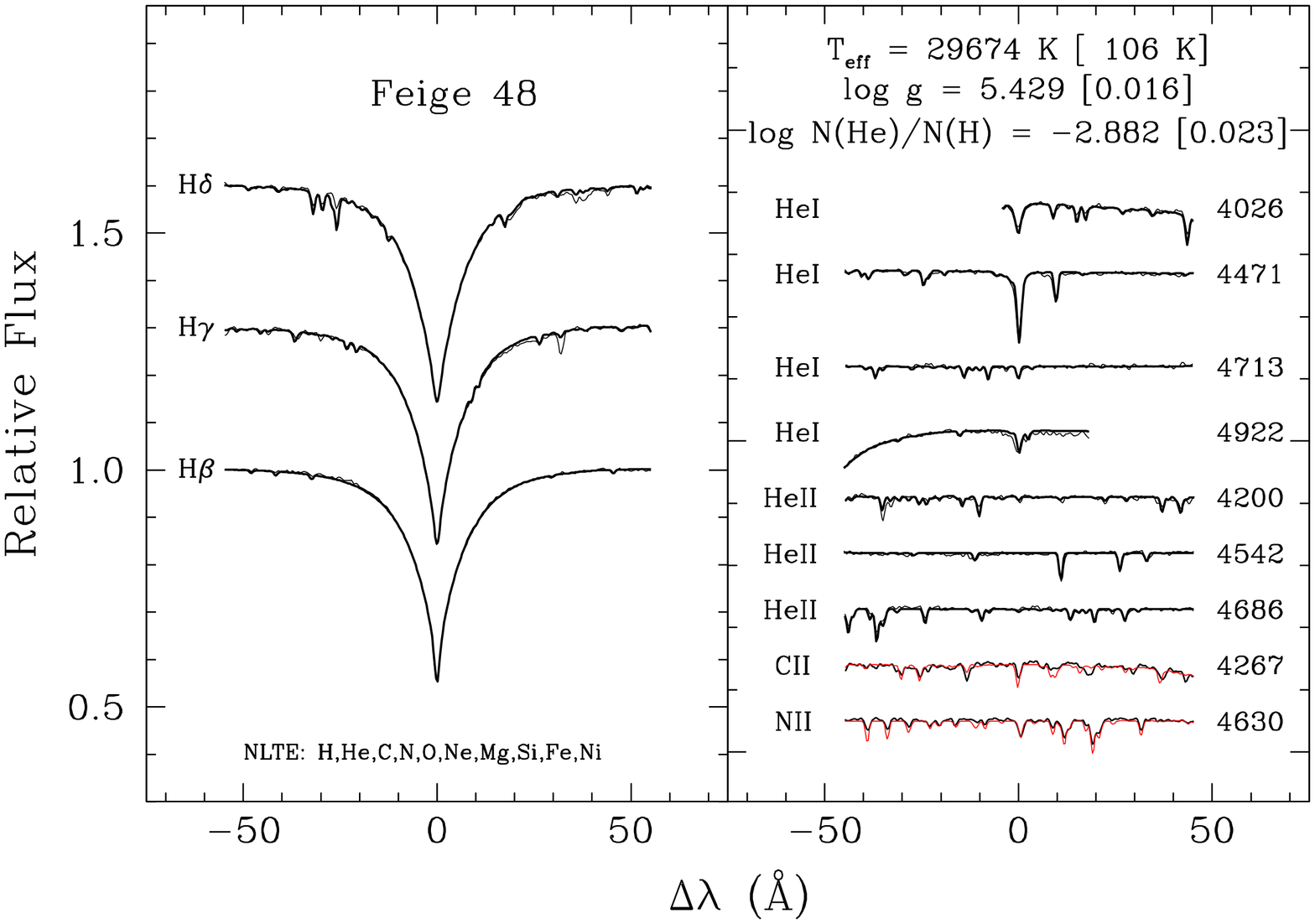}
\includegraphics[scale=0.36,angle=270]{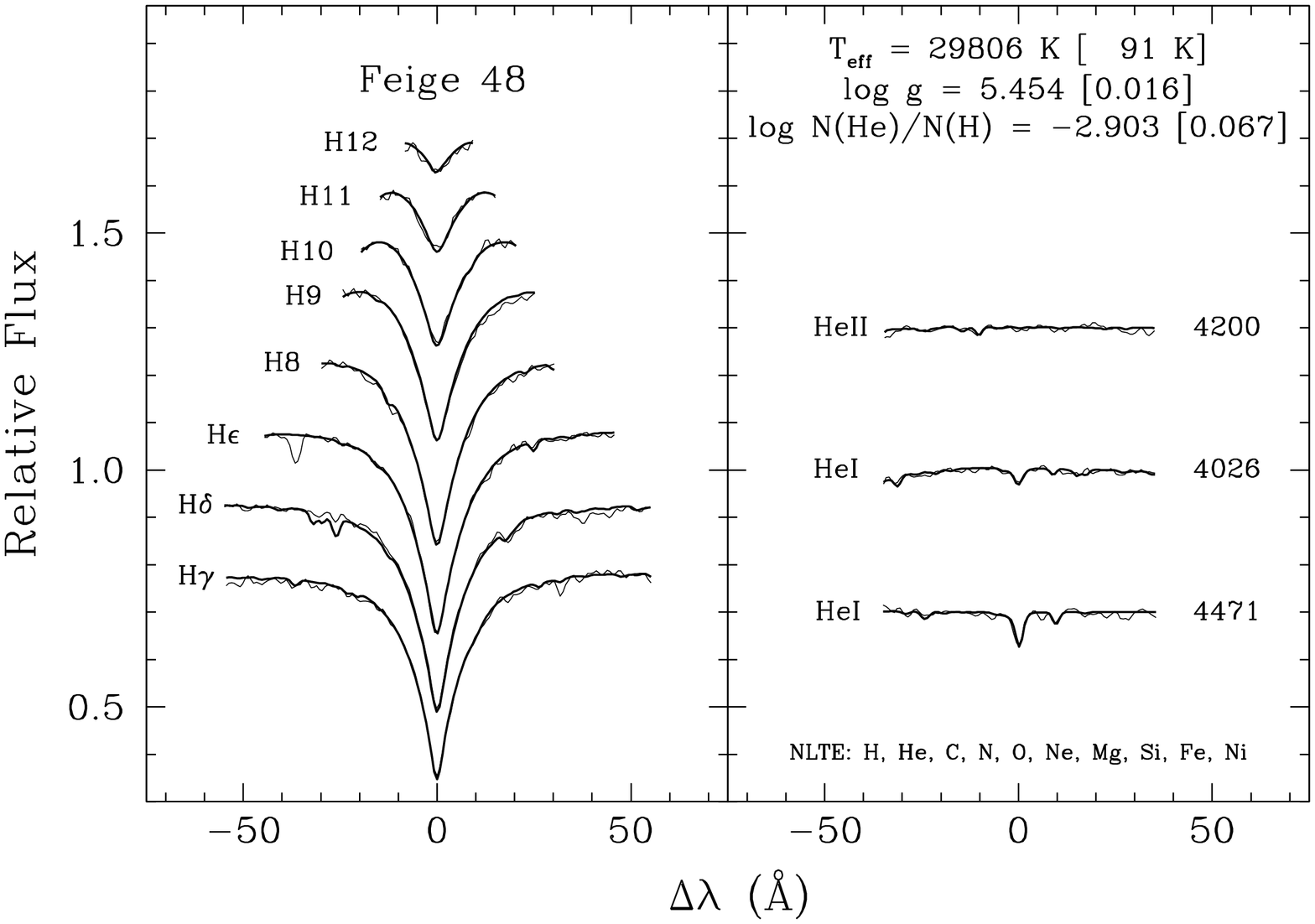}
\includegraphics[scale=0.36,angle=270]{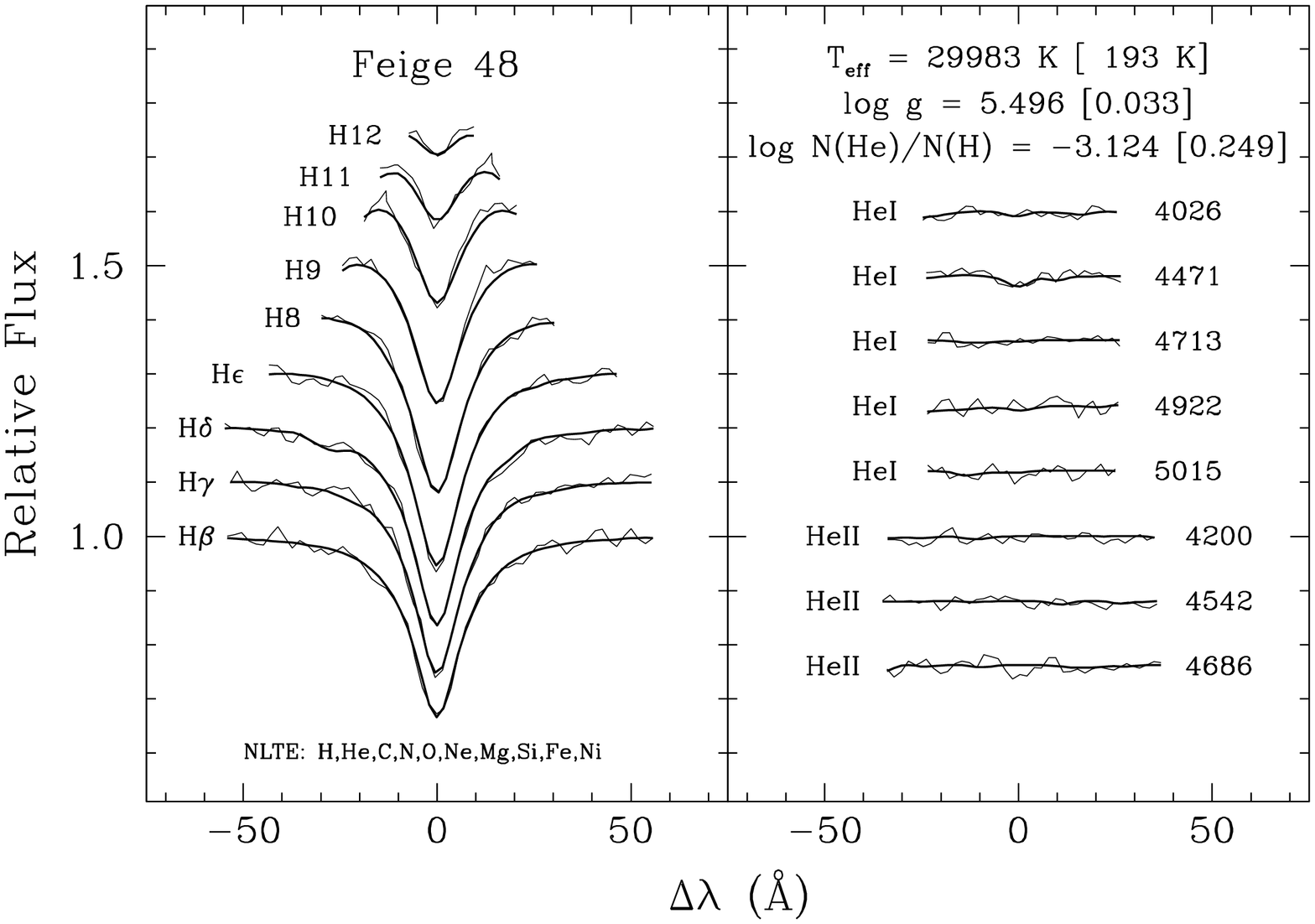}
\includegraphics[scale=0.36,angle=270]{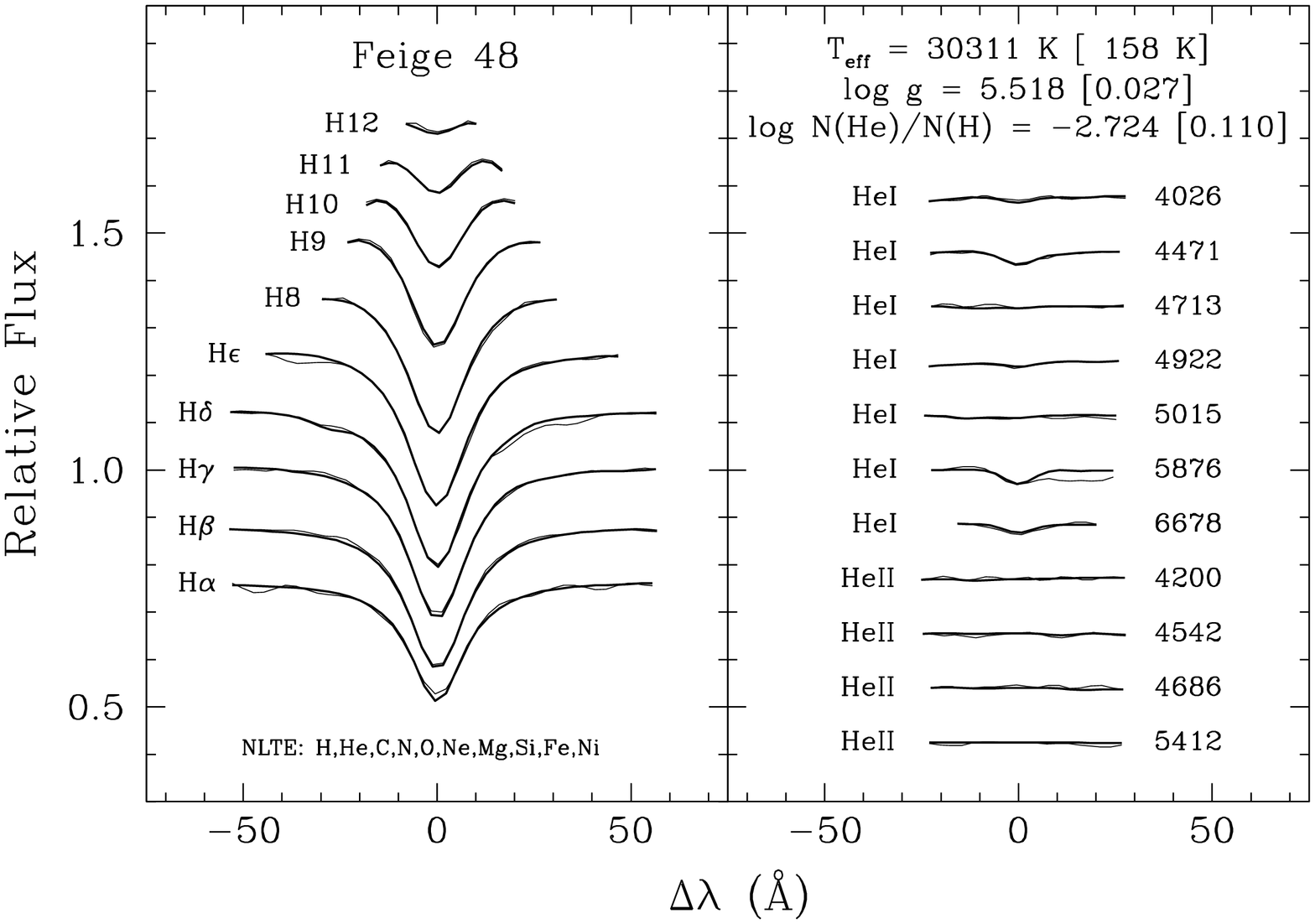}
\caption{Top-left a) Best fit obtained with the 1 \AA~resolution
 MMT spectrum of Feige 48 using our grid of NLTE line-blanketed model
 atmospheres described in Subsection 3.3.2. The Balmer and He~\textsc{i}
 lines are all very well reproduced by the optimal model. 
 Top-right b) Same as (a) but with the bluer 1.9
 \AA~resolution spectrum (BG2).
 Bottom-left c) Same as (a) but with the 6 \AA~resolution
 spectrum (PB6). Note that a sole line of He~\textsc{i} at 4471 \AA~is
 discernible in this spectrum, which explains the lower helium abundance
 found (log \nhe~=$-$3.124) and the larger error associated with this
 value. 
 Bottom-right d) Same as (a) but this time with the lowest
 resolution spectrum (8.7 \AA) BG9. In spite of its low resolution, the
 high S/N allows to distinguish three He~\textsc{i} lines and guess the
 4026 \AA~one.  }
\label{fit}
\end{figure*}

We analyzed the four spectra with both our fully-blanketed model grid and
a more classic metal-free NLTE, H,He grid that was also computed as
indicated just above. We used a $\chi^2$ minimization procedure similar
to that of \citet{saf94}. A simultaneous fit of the Balmer and helium
lines available in our spectra was carried out in order to find the
optimal solution in the three-dimensional space defined by the parameters 
\teff, log $g$, and log \nhe. Prior to that exercise, all the synthetic
spectra (defined by 32001 wavelength points in the range 3500$-$6700
\AA) were degraded by convolution to the experimental resolution of each
of the available spectrum of Feige 48. We note that
microturbulence is a nonissue here given the resolutions of our optical spectra.
The four panels of Figure~\ref{fit} 
show our resulting fits when using the fully-blanketed grid described
in the previous subsection. Panel a) shows the results achieved for the
MMT spectrum, Panel b) for the BG2 spectrum, Panel c) for the PB6
spectrum, and Panel d) for the BG9 spectrum. The results of these fits,
as well as the ones obtained with the H,He grid, are also reported in
Table 3. Note that the quoted uncertainties only reflect the quality of
the fits; they are formal fitting errors.   

When examining the resulting \teff~and log $g$ entries for both grids in
Table 3, one can notice a small systematic trend in the determined values:
from top (MMT) to bottom (BG9), with decreasing resolution, both the
derived effective temperature and the surface gravity increase slightly.
The differences are not large (at most 600 K and $\sim$0.09 dex),
but the trend is nevertheless noticeable. However, the resolution is not
the only difference between the four spectra; the spectral range also
varies and, thus, the lines featured and fitted in each spectrum are not
the same. Hence, the differences in the fitted spectral range could very
well affect the resulting parameters in a systematic way. In order 
to check the effects of varying the spectral range, we carried out some
additional fits with the two lowest resolution spectra. We thus fitted
the BG9 spectrum over the reduced spectral ranges of the MMT, BG2, and PB6
spectra, and the PB6 spectrum over the MMT and BG2 ranges. The results
are shown in Table 3 below the weighted mean values of the four
``conventional'' fits for the two different grids. It should be
mentioned here that our fits usually start at 3740 \AA~(just to the red
of the H12 line), except for the MMT spectrum whose blue limit is at
4000 \AA. 

\begin{deluxetable*}{llccl}[bh!]
\tablewidth{0pt}
\tablecaption{Results of our fitting procedure for Feige 48 }
\tablehead{
\colhead{Spectrum} &
\colhead{\teff} &
\colhead{log $g$} &
\colhead{log \nhe} &
\colhead{Note} \\
\colhead{} &
\colhead{(K)} &
\colhead{(dex)} &
\colhead{(dex)} &
\colhead{}
}
\startdata
\multicolumn{5}{c}{With our NLTE fully line-blanketed model grid} \\
\hline

MMT  &	 29,674 $\pm$ 106 &  5.429 $\pm$ 0.016 & $-$2.882 $\pm$ 0.023 & \\
BG2  &   29,806 $\pm$  91 &  5.454 $\pm$ 0.016 & $-$2.903 $\pm$ 0.067 & \\
PB6  &   29,983 $\pm$ 193 &  5.496 $\pm$ 0.033 & $-$3.124 $\pm$ 0.249 &  \\
BG9  &   30,311 $\pm$ 158 &  5.518 $\pm$ 0.027 & $-$2.724 $\pm$ 0.110 &  \\
Mean &   29,854 $\pm$  60 &  5.459 $\pm$ 0.013 & $-$2.880 $\pm$ 0.021 &  \\
PB6  &   30,010 $\pm$ 169 &  5.487 $\pm$ 0.035 & $-$3.094 $\pm$ 0.256 & with BG2 \\
PB6  &   29,765 $\pm$ 639 &  5.488 $\pm$ 0.093 & $-$3.247 $\pm$ 0.290 & with MMT \\ 
BG9  &   30,264 $\pm$ 113 &  5.517 $\pm$ 0.020 & $-$2.803 $\pm$ 0.101 & with PB6 \\   
BG9  &   30,220 $\pm$ 130 &  5.543 $\pm$ 0.024 & $-$2.836 $\pm$ 0.128 & with BG2 \\
BG9  &   30,483 $\pm$ 384 &  5.521 $\pm$ 0.057 & $-$2.779 $\pm$ 0.099 & with MMT \\

\hline
\multicolumn{5}{c}{With our NLTE H,He model grid} \\
\hline
MMT  & 29,840 $\pm$ 141  & 5.441 $\pm$ 0.024 & $-$2.898 $\pm$ 0.032 & \\ 
BG2  & 30,013 $\pm$  83  & 5.466 $\pm$ 0.016 & $-$2.897 $\pm$ 0.068 & \\
PB6  & 30,161 $\pm$ 167  & 5.502 $\pm$ 0.033 & $-$3.135 $\pm$ 0.259 & \\ 
BG9  & 30,520 $\pm$ 130  & 5.522 $\pm$ 0.027 & $-$2.736 $\pm$ 0.107 & \\
Mean & 30,105 $\pm$  59  & 5.474 $\pm$ 0.011 & $-$2.889 $\pm$ 0.028 & \\
PB6  & 30,182 $\pm$ 173  & 5.502 $\pm$ 0.035 & $-$3.078 $\pm$ 0.258  & with BG2 \\
PB6  & 29,879 $\pm$ 561  & 5.478 $\pm$ 0.094 & $-$3.265 $\pm$ 0.321  & with MMT \\
BG9  & 30,493 $\pm$ 101  & 5.521 $\pm$ 0.021 & $-$2.805 $\pm$ 0.105  & with PB6  \\
BG9  & 30,455 $\pm$ 112  & 5.549 $\pm$ 0.025 & $-$2.847 $\pm$ 0.131  & with BG2 \\
BG9  & 30,694 $\pm$ 340  & 5.537 $\pm$ 0.063 & $-$2.779 $\pm$ 0.111  & with MMT \\

\enddata
\end{deluxetable*}

From Table 3, one can realize that the inferred parameters (effective
temperature, surface gravity, helium abundance) are essentially the same
for the four different 
spectral ranges considered when using the BG9 spectrum. And indeed, the
derived values are the same within the formal fitting errors, and this is
the case for both types of model grids as well. This is true also for
the three different spectral ranges used in conjunction with the PB6
spectrum. We thus conclude that, at low enough resolution, no
significant systematic trend is associated with the choice of the
spectral range. On the other hand, resolution does matter here as can be
seen by comparing the derived parameters obtained with the MMT spectrum,
the PB6 spectrum fitted over the MMT range only, and the BG9 spectrum
again fitted over the MMT range only. Although the formal uncertainties
overlap between the MMT and PB6+MMT cases, there is indeed a small but
significant trend such that the effective temperature and the surface
gravity increase slightly with decreasing resolution as can be inferred
by comparing the MMT and BG9+MMT cases. For its part, the helium
abundance is essentially unchanged as a function of resolution taking
into account the formal errors of the fits. A final look at the effects
of changing the resolution is provided by an experiment in which we
degraded the resolution of the original MMT spectrum through
convolution with a Gaussian with a FWHM of 8.7 \AA. The new inferred
parameters using this degraded spectrum are \teff = 29,922$\pm$100 K,
log $g$ = 5.477$\pm$0.015, and log \nhe = $-$2.805$\pm$0.025, to be
compared with the first line in Table 3 giving \teff~= 29,674$\pm$106 K,
log $g$ = 5.429$\pm$0.016, and log \nhe = $-$2.882$\pm$0.023.

Finally, comparing the atmospheric parameters obtained with the
line-blanketed grid versus the H,He one, we note that adding metals
to NLTE models leads to only slightly lower effective temperatures and
surface gravities. This is exactly what Figure 4 above shows for the
parameters appropriate for Feige 48. As for the effects on the helium
abundance, they are completely negligible. Having done all those fits,
we at last end up with the following atmospheric parameters for Feige 48:
\teff~= 29,854 $\pm$ 60 K, log $g$ = 5.459 $\pm$ 0.013, and log
\nhe~= $-$2.880 $\pm$ 0.021. These are our best estimates, based on the
weighted averages of the values obtained when fitting the four available
spectra with the fully-blanketed grid of model atmospheres (the first
four lines in Table 3). Of course, the quoted uncertainties only reflect
the quality of the fits. 

\begin{figure}[b]
\includegraphics[angle=270,scale=0.41]{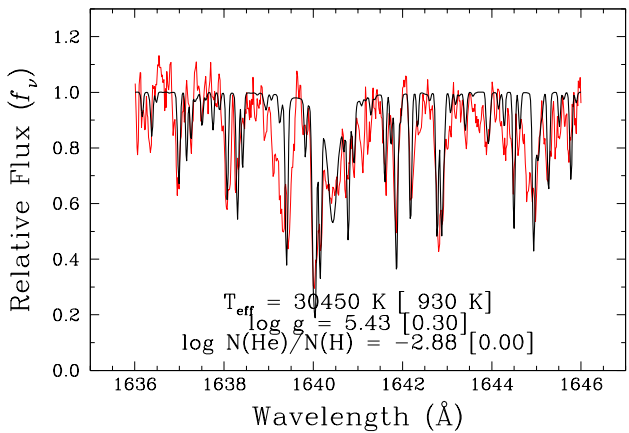}
\caption{Our best fit of the He~\textsc{ii}
$\lambda$1640 line in the STIS spectrum of Feige 48 in terms of
 effective temperature and surface gravity. }
\label{stis}
\end{figure}

A potentially interesting test of our derived atmospheric parameters
for Feige 48 is to attempt fitting the He~\textsc{ii} $\lambda$1640 line
detected in the UV and available in the MAST archives (see Section 3.2). 
This spectral feature is the only one corresponding to the He~\textsc{ii}
ionization stage that is available for that star. Otherwise, there are
no He~\textsc{ii} lines visible in the optical spectrum of Feige 48
because of its relatively low effective temperature and its low helium
abundance as determined just above. Our first attempt to fit the
He~\textsc{ii} $\lambda$1640 line in our standard three-dimensional
search domain (\teff, log $g$ and log \nhe) would not converge to a
unique solution, most likely because of the numerous metallic lines in
the vicinity of the helium line. So we redid the exercise, this time
keeping the helium abundance fixed to the value found previously
($-$2.88), and leaving the fitting procedure find the best match in
terms of effective temperature and gravity only. The program 
did converge this time to a solution giving \teff~= 30,450$\pm$930 K and
log $g$ = 5.43$\pm$0.30, which is perfectly consistent with our previous
results. Even though the fitting errors for the sole He~\textsc{ii}
$\lambda$1640 line are larger than what was obtained with the whole
visible spectra, we take this as a nice consistency check for the
validity of our derived atmospheric parameters. The resulting fit is
presented in Figure~\ref{stis}, where it is possible to see the
He~\textsc{ii} line  blended with Fe~\textsc{iv} and Ni~\textsc{iii}
lines around 1640 \AA~and other Fe~\textsc{iv} lines in its red wing.  

\subsection{Fine Tuning of the Metal Abundances}

Our MMT spectrum, because of its good resolution and excellent
signal-to-noise ratio, features a lot of discernible metallic
lines. When looking closely at our best-fit model shown in Figure 5a, it
can be noticed that the lines of some elements are systematically
stronger or fainter than the observed ones, thus calling for a bit of
fine tuning in order to obtain a better match with the observations. So
we fitted the elements whose lines are visible in the wavelength range
of the MMT spectrum (4020$-$4950 \AA), namely C, N, O, Mg, Si, S and
Fe. These fits were done by using a model atmosphere having the
atmospheric parameters determined by our fit of the MMT spectrum (first
line in Table 3) and the chemical composition mentioned in Section 3.3.1
(with the exception of iron which was not included in the model
atmosphere; see discussion below). Using this model, we computed several
families of synthetic spectra with six or seven different abundances for
each element analyzed. The abundance of the element was changed only in the
computations of the emergent spectrum by SYNSPEC (see the SYNSPEC user's
guide for more details\footnote{http://nova.astro.umd.edu/Tlusty2002/tlusty-frames-guides.html}). 
This approach is not entirely self-consistent, but because the spectrum of 
Feige 48 turns out to be not much affected by line-blanketing nor NLTE 
effects, and also because we do not expect drastic changes in the 
abundances, we think this method is reasonable for this particular case.

The detailed comparison of the observed MMT spectrum with our final best
synthetic one (having adjusted metal abundances) is shown in
Figure~\ref{fitmetal}. 
The original best-fit spectrum, seen in Figure \ref{fit}a, is also featured
in the comparison (dotted line). This way the improvement brought by the 
fine tuning can easily be seen. The new abundances obtained by fitting 
the metallic lines of the MMT spectra are indicated in Table 1. 
The uncertainties were obtained from the standard deviation of the 
abundances indicated by different lines of a same element, or by eye when 
only one line of the element was fitted.
Specifically, we decreased by a few
tenths of a dex the abundances of carbon, nitrogen, and oxygen. This resulted in a better agreement between the
observed and synthetic spectrum for most of the lines originating from 
O~\textsc{ii}, N~\textsc{ii}, and the C~\textsc{ii} doublet at 4267~\AA.
The only noteworthy remaining
discrepancies (after our adjustments) for these elements are the
O~\textsc{ii} lines at 4075.8 \AA~that are still too strong. Magnesium
and silicon kept roughly the same abundances we considered initially;
in those cases, their lines were already well reproduced with our
best-fit model. 
Sulfur was not included in the model
atmospheres at the outset, but when added into the synthetic spectrum,
its lines, including one blended with an O~\textsc{ii} line at our
resolution, were better reproduced (see Figure 7b) with an abundance a bit higher than the mean one indicated in Table 1.

\begin{figure*}[p]
\includegraphics[angle=90,scale=0.75]{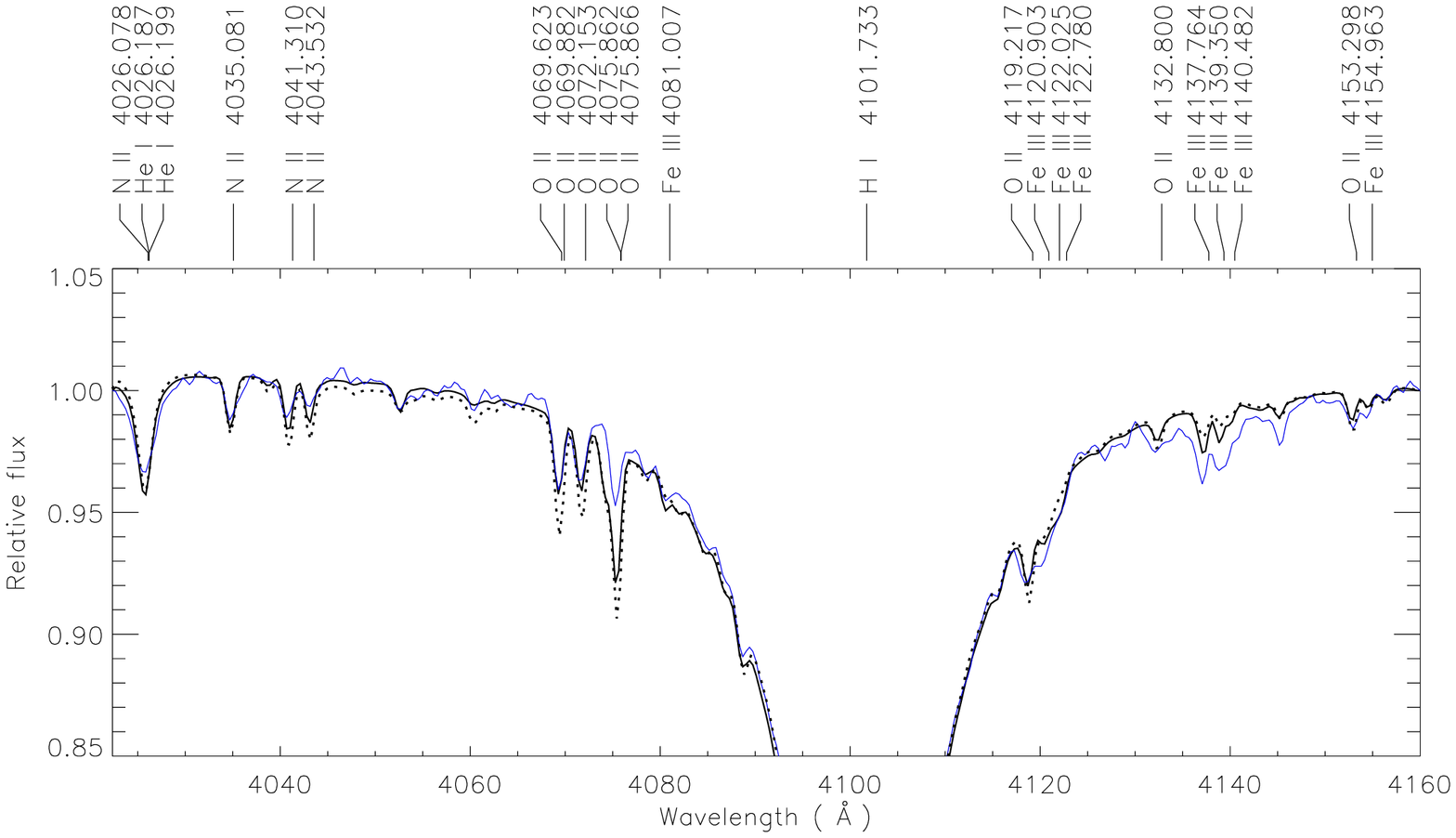}
\includegraphics[angle=90,scale=0.75]{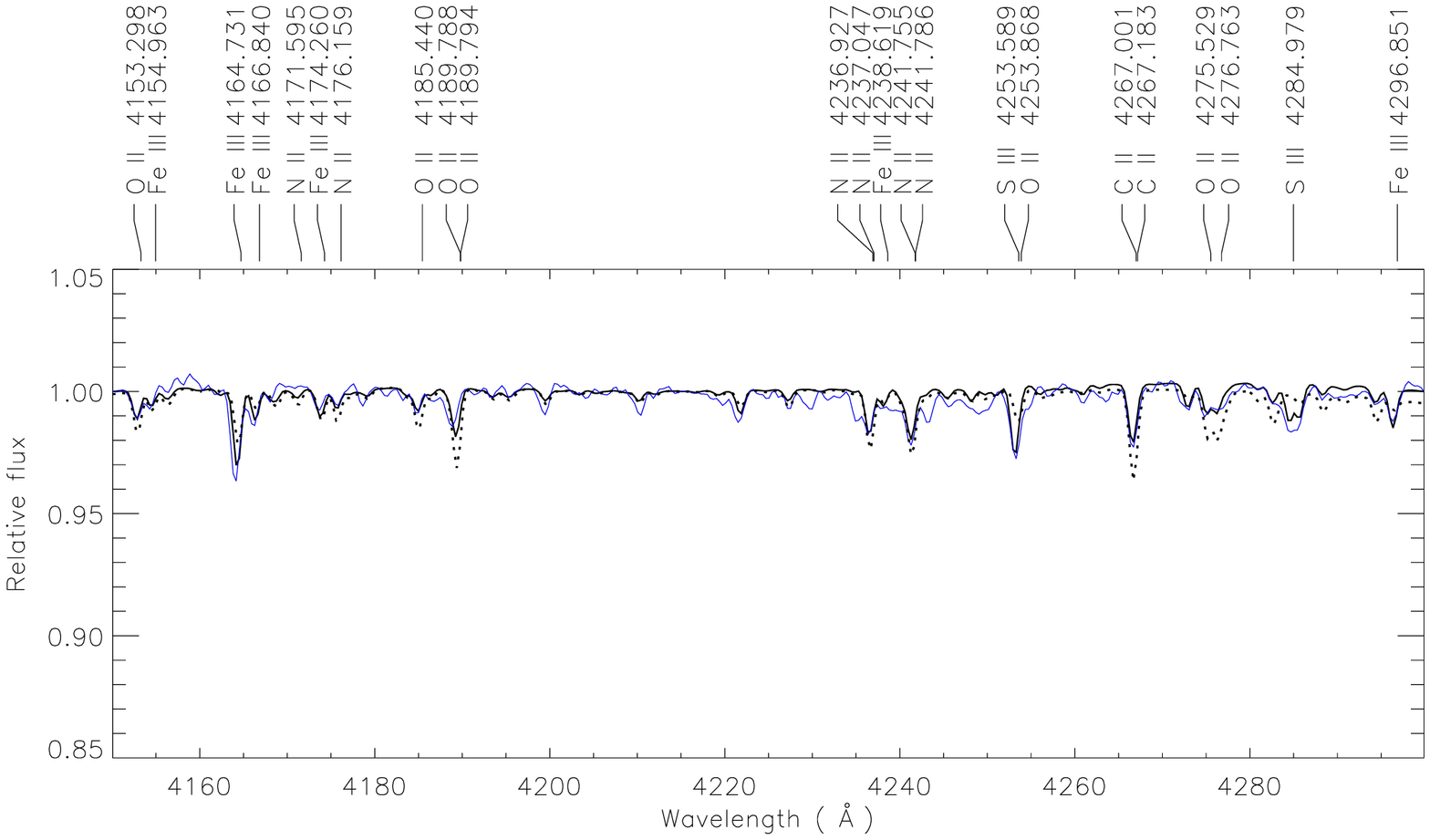}
 \caption{Detailed comparison of the observed
MMT spectrum (blue line) with a synthetic spectrum (black line) having
the abundances fine tuned in order to obtain a better match.
The model atmosphere used for generating the synthetic spectrum has the parameters
found by the fitting procedure of the MMT spectrum : \teff~= 29,674 K,
log $g$ = 5.43 and log \nhe~= $-$2.88. The synthetic spectra obtained from
the fitting procedure (in Fig \ref{fit}a) is also shown (dotted line).
The main absorption features are
indicated with the name of the ion and the wavelength of the transition.  
}
 \label{fitmetal}
\end{figure*}

\addtocounter{figure}{-1}
 \begin{figure*}[p]
\includegraphics[angle=90,scale=0.75]{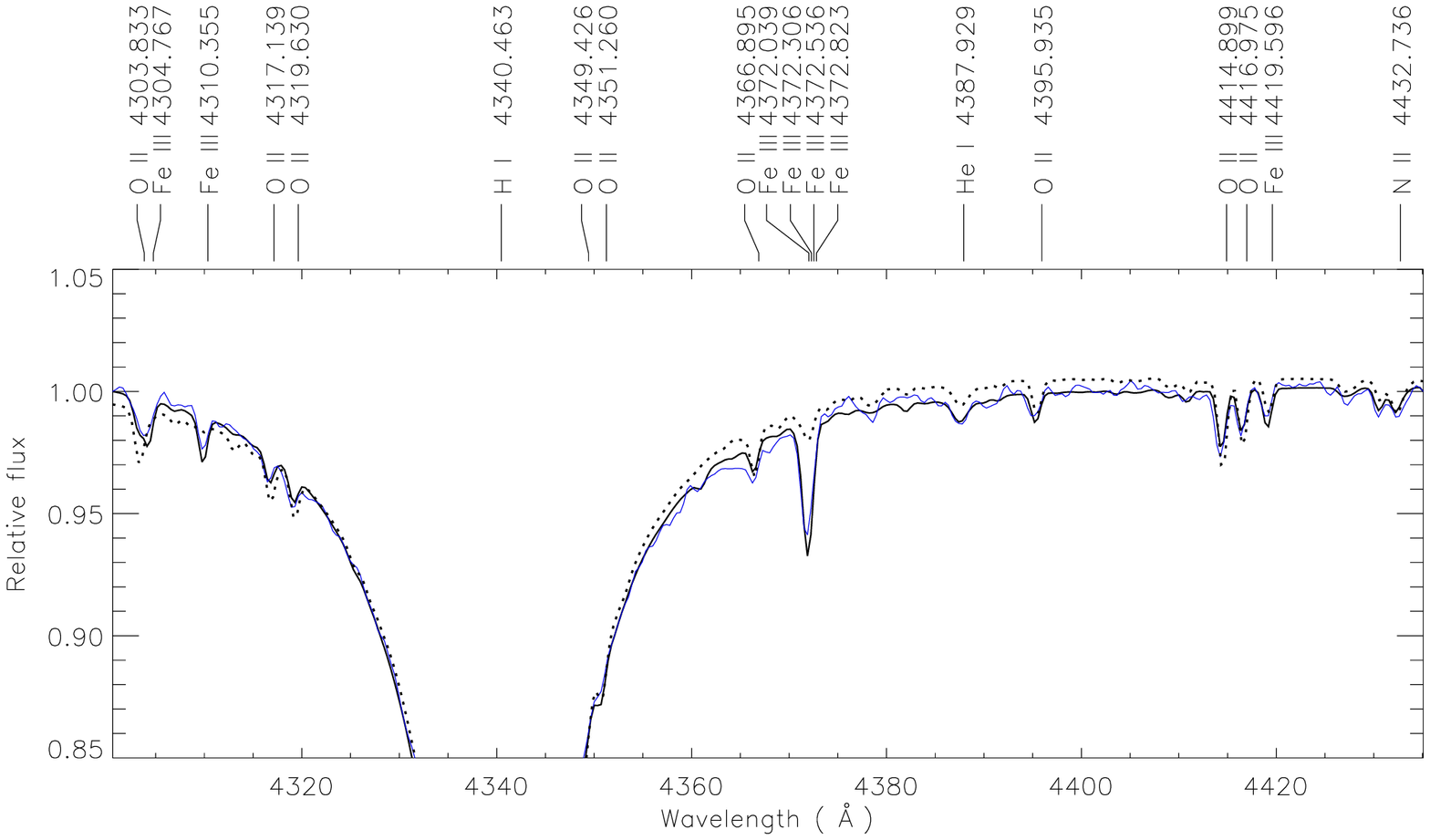}
\includegraphics[angle=90,scale=0.75]{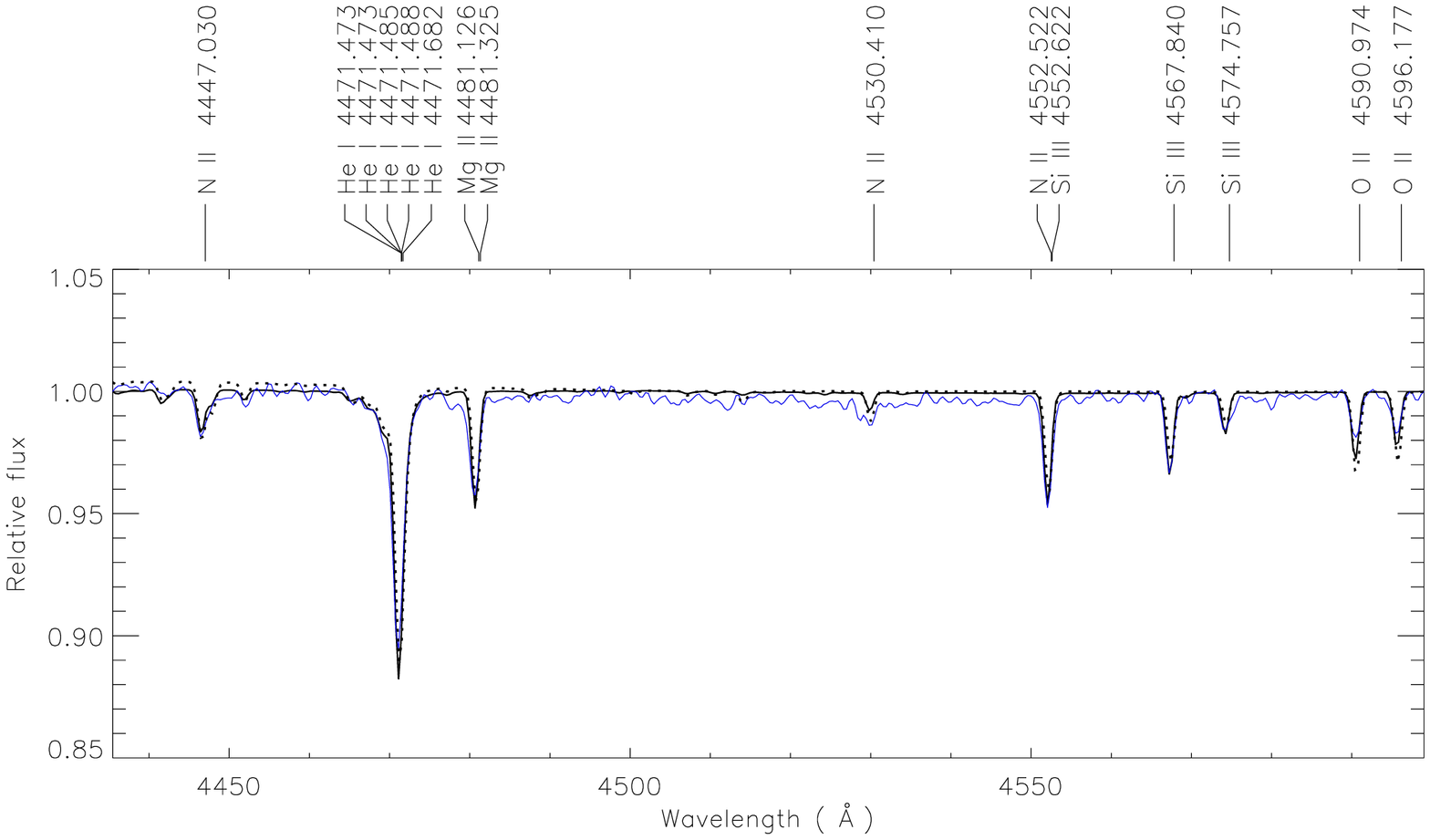}
 \caption{Continued. }
\end{figure*}

\addtocounter{figure}{-1}
 \begin{figure*}[p]
\includegraphics[angle=90,scale=0.75]{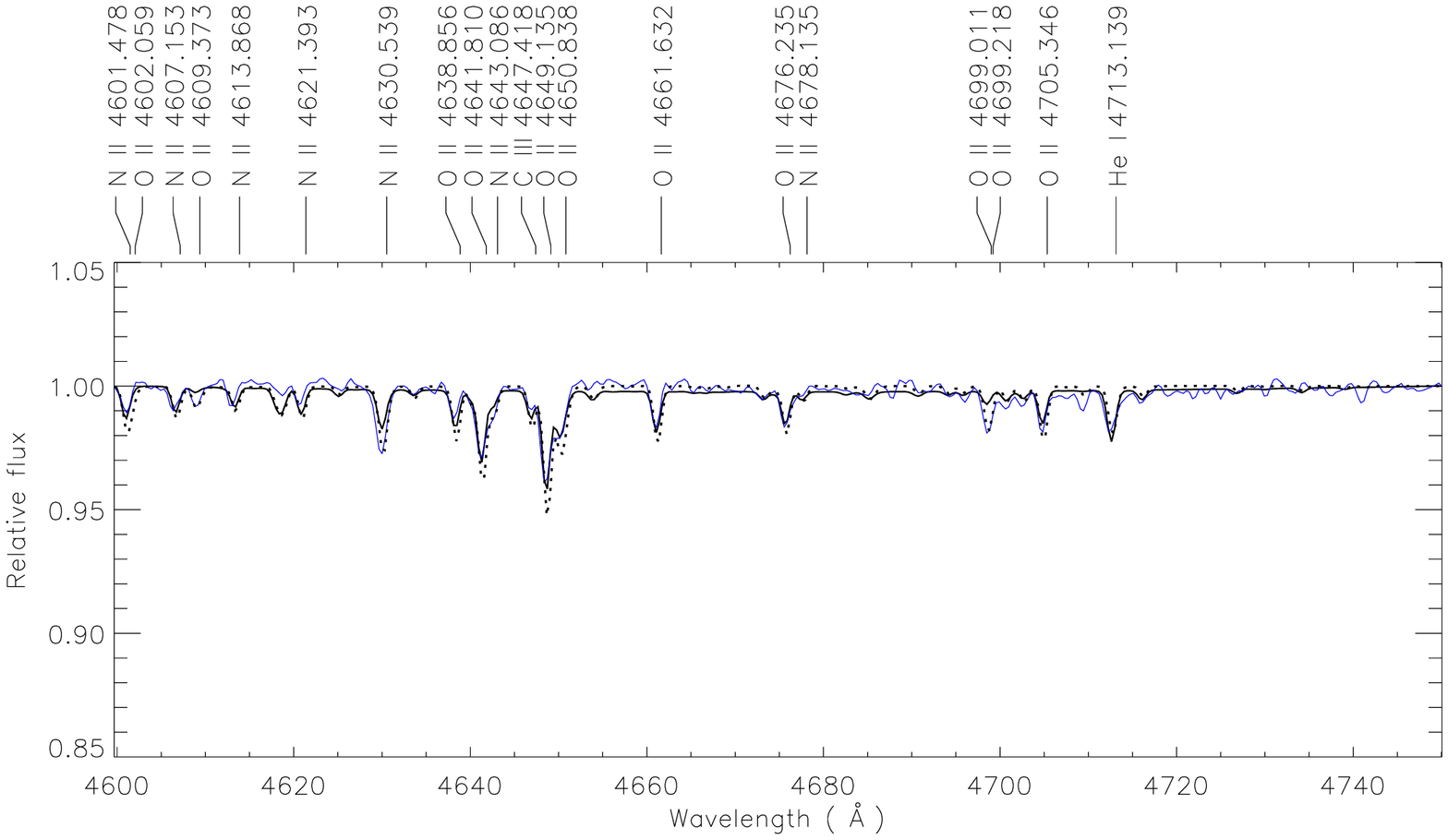}
\includegraphics[angle=90,scale=0.75]{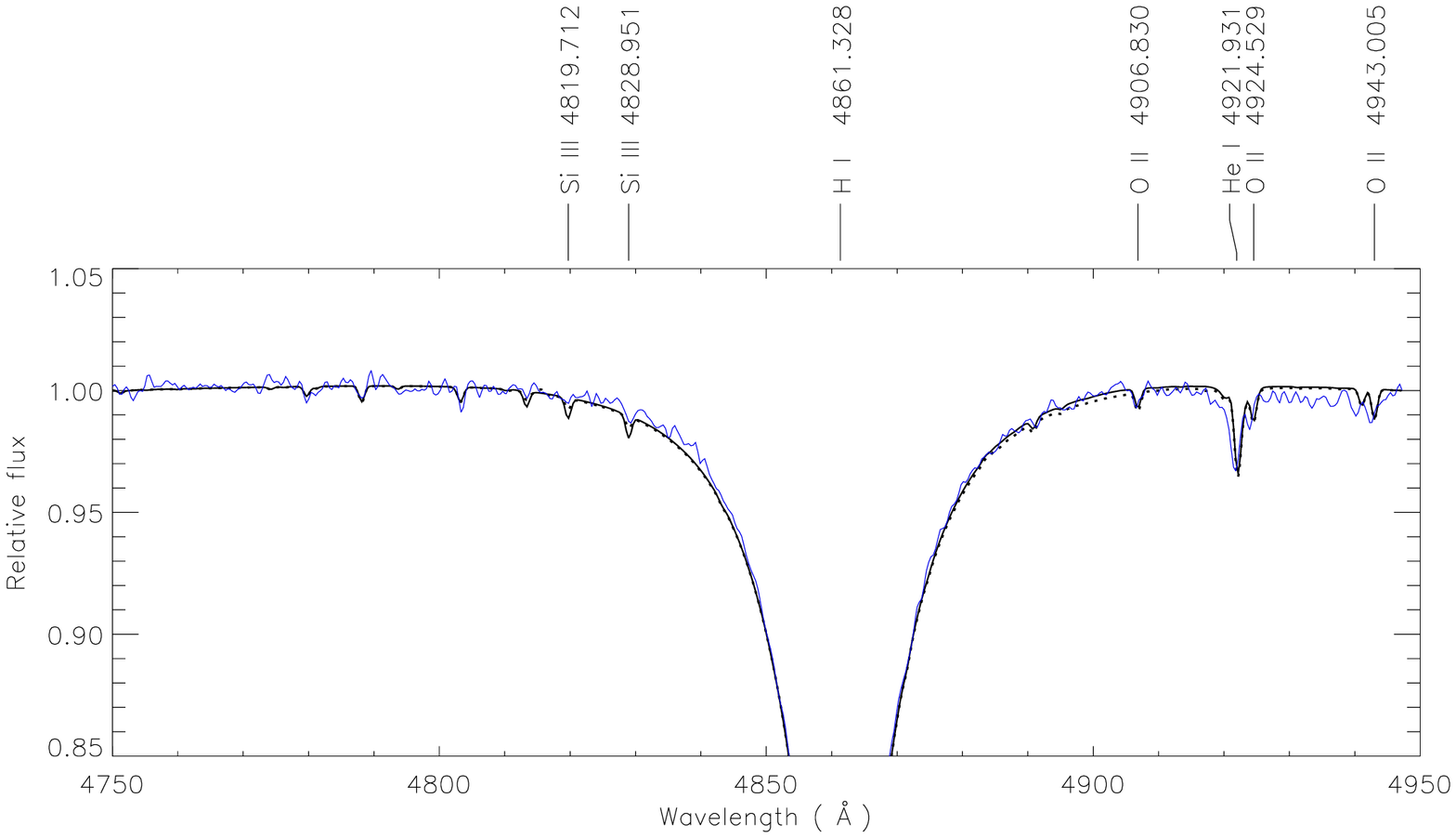}
 \caption{Continued. There is an offset of the
 wavelengths from He~\textsc{i} $\lambda$4922 until the end of the
 spectrum, probably due to the wavelength calibration that is not exactly
 right at the very end of the spectrum. }
\end{figure*}

Finally, iron was a particular case. We noticed that most of the iron
lines were too shallow in our optimal synthetic spectrum, and we were
unable to obtain a good match when changing the abundance in the
synthetic spectra or in the model atmosphere directly. After a thorough
inspection of the synthetic spectra, it appears that some emission was
occuring in a few iron lines, an unexpected and certainly unrealistic
result. The two most problematic sets of lines (the complex around 4372
\AA~and Fe~\textsc{iii} $\lambda$4310.355) originate from within two
specific superlevels, and we suspect that there might be a problem with
the populations of the different components of the superlevels. These
components (within a given superlevel) are assumed to be in Boltzmann
equilibrium with respect to each other. The solution we found to avoid this
problem was to add iron in a synthetic spectrum computed from a model
atmosphere that does not include iron. In this way, SYNSPEC computes the
iron population in LTE and no strange emission occurs. When using these
kind of spectra, we found a good match to the observed lines
corresponding to an optimal iron abundance of log
$N$(Fe)/$N$(H) = $-$4.48, which is very near the mean value obtained in
Table 1.    

Overall we can conclude that only minor adjustments were required in 
order to match in a better way the metallic lines featured in our MMT
spectrum.Our new abundances for the seven species seen in this spectrum 
all agree very well with the previous measurements made in the past studies
reported in Table 1.

\subsection{Interstellar Reddening}

The $\chi^2$ method that we have used to derive the atmospheric parameters of Feige 48
is not sensitive to the presence of interstellar reddening since it relies on the
(simultaneous) fits of the line profiles of hydrogen and helium, and not on the
overall flux-calibrated spectrum. Nevertheless, it is instructive to verify after
the fact if reddening is indeed small, as assumed by construction in our $\chi^2$
approach.

From our best-fitting synthetic spectrum, characterized by \teff\ = 29,850 K, log $g$ = 5.46, log N(He)/N(H) = -2.88, and the metallicity given in the 7$^{th}$ column of
Table 1, we derive an unreddened color index $(B-V)_0$ = $-$0.255. In comparison, the
best available UBV photometry on Feige 48, that of \citet{bern73}, gives
$V$ = 13.480 $\pm$ 0.026, $B-V$ = $-$0.250 $\pm$ 0.013, and $U-B$ = $-$1.030 $\pm$ 0.017.
Neglecting the uncertainties on the model color index, this immediately leads to a
reddening index of $E(B-V)$ = 0.005 $\pm$ 0.013, which is indeed quite low.

An independent check on that result is provided by the use of the Str\"{o}mgren
photometry of \citet{wes92} which gives, for Feige 48, $V$ = $y$ = 13.456 $\pm$
0.029, $b-y$ = $-$0.116 $\pm$ 0.017, $u-b$ = $-$0.162 $\pm$ 0.018, and $m_1$ = 0.086 $\pm$ 0.018.
With the help of the relation $E(b-y)$ = 0.642 $E(B-V)$ -- which can be derived from the
equations given in \citet{sea79} -- and the unreddened model value of $(b-y)_0$ =
$-$0.123, we find a reddening index of $E(B-V)$ = 0.011 $\pm$ 0.036, less accurate but entirely compatible with our first estimate.

\section{CONCLUSION}

As part of an ongoing major effort to exploit fully the asteroseismic
potential of Feige 48 (a rare bright rapid pulsator part of a close
binary system), our work aimed at obtaining the most accurate atmospheric 
parameters possible for this sdB star. To achieve that goal, we analyzed
four time-averaged optical spectra of this star --- three of them having
exceptionally high S/N --- with state-of-the-art NLTE line-blanketed
model atmospheres including the eight most abundant metallic species
observed in the star's atmosphere, namely, C, N, O, Ne, Mg, Si, Fe, and
Ni. In comparison, previous atmospheric studies of Feige 48 were based
on either line-blanketed LTE models with arbitrary metallicities or
metal-free LTE and NLTE models. 

Our final adopted parameters are the weighted mean of the solutions
found with the different spectra: \teff = 29,850 $\pm$ 60 K, log $g$ =
5.46 $\pm$ 0.01, and log \nhe = $-$2.88 $\pm$ 0.02, with the quoted
uncertainties measuring only the quality of the spectral fits. A similar
analysis made, this time, with more classical metal-free, H,He NLTE
models led to very similar atmospheric parameters as indicated in Table
3. This demonstrates that the effects of combining the NLTE approach
with metal line blanketing are not very large in the atmosphere of Feige
48. As depicted in Figure \ref{map}, part of this is accidental and
specific to this particular star. Presumably, this coincidence explains
why our derived parameters agree rather well with previous estimates
based on less sophisticated models, implying that the current estimations 
of the atmospheric parameters of Feige 48 are reliable and sound. This is a good thing, 
particularly from an asteroseismological point of view, because it means that the
spectroscopic constraints to be used in seismic studies, including past
efforts, can be trusted.

During our investigations, we noticed a slight, but possibly significant
systematic trend suggesting an increase of both the derived effective
temperature and the surface gravity with decreasing resolution (from 1.0
to 8.7 \AA~ in our data), while the helium abundance appeared to be
insensitive. The overall differences are relatively small, $\sim$600
K and $\sim$0.09 dex, but the trend is seen in the fits carried out with
both the metal-free and the line-blanketed grids. This is the first time we
observe this effect in a sdB star, possibly because we had at our
disposal exceptionally high S/N spectra.
\citet{Geier13b} investigated the differences in effective parameters 
derived from medium resolution spectra and high-resolution echelle ones, they
found the averages of the shifts to be 
$\Delta$\teff $\simeq$ 1100 K and $\Delta$ log $g$ $\simeq$ 0.12 and no systematic
trend was seen. The differences we found for Feige 48 are consistent with these values.
We also found that for
the lower resolution spectra (PB6 and BG9), changing the spectral range
of the fit leads to, within the fitting errors, the same estimates of
the derived parameters and, thus, no observable systematic effects.

We also inspected the effects that the various metallic elements considered
in the NLTE models have on the atmospheric structure. In the specific
case of the models computed for Feige 48, the temperature in the
outermost atmospheric layers is highly sensitive to the presence of the
light metals, while the influence of iron and  nickel is mostly confined
to layers deeper than log $m$ $\simeq$ $-$3.0. The actual amount of
nickel present in the atmosphere of the star, log $N$(Ni)/$N$(H) =
$-$5.31, does not produce a significant added effect on the temperature
structure modified already by the presence of the light metals and
iron.

As a consistency check, we fitted the only He \textsc{ii} spectral
feature that we could find in the spectrum of Feige 48 which, otherwise, 
contains only H \textsc{i} and He \textsc{i} lines. And indeed,
through the MAST archives, we retrieved and fitted the weak He
\textsc{ii} line at 1640 \AA~present in the STIS spectrum of Feige 48.
This was done in terms of effective temperature and surface gravity only
(the helium abundance being fixed), leading to the independent
estimates of \teff~= 30,450$\pm$930 K and log $g$ = 5.43$\pm$0.30, fully
consistent with our values derived from the optical data. 

Our best data set was the MMT spectrum, characterized by the relatively
high resolution of 1.0 \AA~ over a spectral range 4000$-$4950 \AA, and
the very high value of S/N $\simeq$ 460. This particular spectrum shows
a host of distinct metallic lines that could be examined in details. 
With the initial metallicity specified as in Table 1, a good agreement
was obtained between the observed and predicted metal lines as can be
seen in Figure \ref{fit}a. However, a distinct improvement was reached
when slightly adjusting the individual metal abundances as was done in
Subsection 3.5. The detailed results are presented in the series of
Figure 7. 

The atmospheric parameters derived here for Feige 48 will provide an
essential ingredient in the upcoming new seismic analysis of that star,
which will be based on the recent extensive photometric campaign
that has revealed some 46 pulsation modes compared to the 9 modes
previously known (Green et al., in preparation). Likewise, these
estimates form the basis of the method that is used to exploit the
signature that the degree index $\ell$ of a pulsation mode leaves on the
wavelength-amplitude relationship (see, e.g., \citealt{rand2005}). In the
specific case of Feige 48, \citet{fon2006} and \citet{quirion2010} have
presented preliminary efforts to exploit this signature by comparing
optical with FUV amplitudes, and these certainly deserves to be pushed
further on the basis of our improved determinations of the atmospheric
parameters of that pulsator. A priori mode identification can be
extremely useful in the search for an optimal seismic model in parameter
space \citep{vang08}. 

\acknowledgments

This work was supported in part by the Natural Sciences and Engineering
Research Council of Canada through a doctoral fellowship awarded to M.L. and
through a research grant awarded to G.F. The latter also acknowledges
the contribution of the Canada Research Chair Program. We are also most
grateful to Pierre Bergeron for providing us with a spectrum of Feige 48.

\bibliographystyle{apj}
\bibliography{referencef48}

\begin{thebibliography}{40}
\expandafter\ifx\csname natexlab\endcsname\relax\def\natexlab#1{#1}\fi

\bibitem[{{Bern} \& {Wramdemark}(1973)}]{bern73}
{Bern}, K. \& {Wramdemark}, S. 1973, Lowell Observatory Bulletin, 8, 1

\bibitem[{{Blanchette} {et~al.}(2008){Blanchette}, {Chayer}, {Wesemael},
  {Fontaine}, {Fontaine}, {Dupuis}, {Kruk}, \& {Green}}]{blan2008}
{Blanchette}, J.-P., {Chayer}, P., {Wesemael}, F., {Fontaine}, G., {Fontaine},
  M., {Dupuis}, J., {Kruk}, J.~W., \& {Green}, E.~M. 2008, \apj, 678, 1329

\bibitem[{{Brassard} {et~al.}(2001){Brassard}, {Fontaine}, {Bill{\`e}res},
  {Charpinet}, {Liebert}, \& {Saffer}}]{brassard2001}
{Brassard}, P., {Fontaine}, G., {Bill{\`e}res}, M., {Charpinet}, S., {Liebert},
  J., \& {Saffer}, R.~A. 2001, \apj, 563, 1013

\bibitem[{{Brassard} {et~al.}(2010){Brassard}, {Fontaine}, {Chayer}, \&
  {Green}}]{bra10}
{Brassard}, P., {Fontaine}, G., {Chayer}, P., \& {Green}, E.~M. 2010, in
  American Institute of Physics Conference Series, Vol. 1273, American
  Institute of Physics Conference Series, ed. K.~{Werner} \& T.~{Rauch},
  259--262

\bibitem[{{Charpinet} {et~al.}(2005{\natexlab{a}}){Charpinet}, {Fontaine},
  {Brassard}, {Bill{\`e}res}, {Green}, \& {Chayer}}]{char05f48}
{Charpinet}, S., {Fontaine}, G., {Brassard}, P., {Bill{\`e}res}, M., {Green},
  E.~M., \& {Chayer}, P. 2005{\natexlab{a}}, \aap, 443, 251

\bibitem[{{Charpinet} {et~al.}(1997){Charpinet}, {Fontaine}, {Brassard},
  {Chayer}, {Rogers}, {Iglesias}, \& {Dorman}}]{char1997}
{Charpinet}, S., {Fontaine}, G., {Brassard}, P., {Chayer}, P., {Rogers}, F.~J.,
  {Iglesias}, C.~A., \& {Dorman}, B. 1997, \apjl, 483, L123

\bibitem[{{Charpinet} {et~al.}(1996){Charpinet}, {Fontaine}, {Brassard}, \&
  {Dorman}}]{char1996}
{Charpinet}, S., {Fontaine}, G., {Brassard}, P., \& {Dorman}, B. 1996, \apjl,
  471, L103

\bibitem[{{Charpinet} {et~al.}(2005{\natexlab{b}}){Charpinet}, {Fontaine},
  {Brassard}, {Green}, \& {Chayer}}]{char05pg}
{Charpinet}, S., {Fontaine}, G., {Brassard}, P., {Green}, E.~M., \& {Chayer},
  P. 2005{\natexlab{b}}, \aap, 437, 575

\bibitem[{{Charpinet} {et~al.}(2013){Charpinet}, {Van Grootel}, {Brassard},
  {Fontaine}, {Green}, \& {Randall}}]{charp2013}
{Charpinet}, S., {Van Grootel}, V., {Brassard}, P., {Fontaine}, G., {Green},
  E.~M., \& {Randall}, S.~K. 2013, in European Physical Journal Web of
  Conferences, Vol.~43, European Physical Journal Web of Conferences, 4005

\bibitem[{{Chayer} {et~al.}(2004){Chayer}, {Fontaine}, {Fontaine},
  {Lamontagne}, {Wesemael}, {Dupuis}, {Heber}, {Napiwotzki}, \&
  {Moehler}}]{chay04}
{Chayer}, P., {Fontaine}, G., {Fontaine}, M., {Lamontagne}, R., {Wesemael}, F.,
  {Dupuis}, J., {Heber}, U., {Napiwotzki}, R., \& {Moehler}, S. 2004, \apss,
  291, 359

\bibitem[{{Fontaine} {et~al.}(2003){Fontaine}, {Brassard}, {Charpinet},
  {Green}, {Chayer}, {Bill{\`e}res}, \& {Randall}}]{fon2003}
{Fontaine}, G., {Brassard}, P., {Charpinet}, S., {Green}, E.~M., {Chayer}, P.,
  {Bill{\`e}res}, M., \& {Randall}, S.~K. 2003, \apj, 597, 518

\bibitem[{{Fontaine} \& {Chayer}(2006)}]{fon2006}
{Fontaine}, G. \& {Chayer}, P. 2006, in Astronomical Society of the Pacific
  Conference Series, Vol. 348, Astrophysics in the Far Ultraviolet: Five Years
  of Discovery with FUSE, ed. G.~{Sonneborn}, H.~W. {Moos}, \& B.-G.
  {Andersson}, 181

\bibitem[{{Geier}(2013)}]{geier13}
{Geier}, S. 2013, \aap, 549, A110

\bibitem[{{Geier} {et~al.}(2013){Geier}, {Heber}, {Edelmann}, {Morales-Rueda},
  {Kilkenny}, {O'Donoghue}, {Marsh}, \& {Copperwheat}}]{Geier13b}
{Geier}, S., {Heber}, U., {Edelmann}, H., {Morales-Rueda}, L., {Kilkenny}, D.,
  {O'Donoghue}, D., {Marsh}, T.~R., \& {Copperwheat}, C. 2013, \aap, 557, A122

\bibitem[{{Green} {et~al.}(2003){Green}, {Fontaine}, {Reed}, {Callerame},
  {Seitenzahl}, {White}, {Hyde}, {{\O}stensen}, {Cordes}, {Brassard}, {Falter},
  {Jeffery}, {Dreizler}, {Schuh}, {Giovanni}, {Jeffery}, {Dreizler}, {Schuh},
  {Giovanni}, {Edelmann}, {Rigby}, \& {Bronowska}}]{green2003}
{Green}, E.~M., {Fontaine}, G., {Reed}, M.~D., {Callerame}, K., {Seitenzahl},
  I.~R., {White}, B.~A., {Hyde}, E.~A., {{\O}stensen}, R., {Cordes}, O.,
  {Brassard}, P., {Falter}, S., {Jeffery}, E.~J., {Dreizler}, S., {Schuh},
  S.~L., {Giovanni}, M., {Jeffery}, E.~J., {Dreizler}, S., {Schuh}, S.~L.,
  {Giovanni}, M., {Edelmann}, H., {Rigby}, J., \& {Bronowska}, A. 2003, \apjl,
  583, L31

\bibitem[{{Heber}(2009)}]{heber2009}
{Heber}, U. 2009, \araa, 47, 211

\bibitem[{{Heber} {et~al.}(2000){Heber}, {Reid}, \& {Werner}}]{heb00}
{Heber}, U., {Reid}, I.~N., \& {Werner}, K. 2000, \aap, 363, 198

\bibitem[{{Kilkenny} {et~al.}(2010){Kilkenny}, {Fontaine}, {Green}, \&
  {Schuh}}]{kil2010}
{Kilkenny}, D., {Fontaine}, G., {Green}, E.~M., \& {Schuh}, S. 2010,
  Information Bulletin on Variable Stars, 5927, 1

\bibitem[{{Kilkenny} {et~al.}(1997){Kilkenny}, {Koen}, {O'Donoghue}, \&
  {Stobie}}]{kilkenny1997}
{Kilkenny}, D., {Koen}, C., {O'Donoghue}, D., \& {Stobie}, R.~S. 1997, \mnras,
  285, 640

\bibitem[{{Koen} {et~al.}(1997){Koen}, {Kilkenny}, {O'Donoghue}, {van Wyk}, \&
  {Stobie}}]{koen1997}
{Koen}, C., {Kilkenny}, D., {O'Donoghue}, D., {van Wyk}, F., \& {Stobie}, R.~S.
  1997, \mnras, 285, 645

\bibitem[{{Koen} {et~al.}(1998){Koen}, {O'Donoghue}, {Pollacco}, \&
  {Nitta}}]{koen98}
{Koen}, C., {O'Donoghue}, D., {Pollacco}, D.~L., \& {Nitta}, A. 1998, \mnras,
  300, 1105

\bibitem[{{Lanz} \& {Hubeny}(1995)}]{lanz95}
{Lanz}, T. \& {Hubeny}, I. 1995, \apj, 439, 905

\bibitem[{{Lanz} \& {Hubeny}(2003{\natexlab{a}})}]{lanz03}
---. 2003{\natexlab{a}}, \apjs, 146, 417

\bibitem[{{Lanz} \& {Hubeny}(2003{\natexlab{b}})}]{lanz03atom}
{Lanz}, T. \& {Hubeny}, I. 2003{\natexlab{b}}, in Astronomical Society of the
  Pacific Conference Series, Vol. 288, Stellar Atmosphere Modeling, ed.
  I.~{Hubeny}, D.~{Mihalas}, \& K.~{Werner}, 117

\bibitem[{{Lanz} \& {Hubeny}(2007)}]{lanz07}
---. 2007, \apjs, 169, 83

\bibitem[{{Latour} {et~al.}(2011){Latour}, {Fontaine}, {Brassard}, {Green},
  {Chayer}, \& {Randall}}]{lat11}
{Latour}, M., {Fontaine}, G., {Brassard}, P., {Green}, E.~M., {Chayer}, P., \&
  {Randall}, S.~K. 2011, \apj, 733, 100

\bibitem[{{Latour} {et~al.}(2013){Latour}, {Fontaine}, {Chayer}, \&
  {Brassard}}]{lat2013}
{Latour}, M., {Fontaine}, G., {Chayer}, P., \& {Brassard}, P. 2013, \apj, 773,
  84

\bibitem[{{Napiwotzki}(1997)}]{nap97}
{Napiwotzki}, R. 1997, \aap, 322, 256

\bibitem[{{N{\'e}meth} {et~al.}(2012){N{\'e}meth}, {Kawka}, \&
  {Vennes}}]{nemeth2012}
{N{\'e}meth}, P., {Kawka}, A., \& {Vennes}, S. 2012, \mnras, 427, 2180

\bibitem[{{O'Donoghue} {et~al.}(1997){O'Donoghue}, {Lynas-Gray}, {Kilkenny},
  {Stobie}, \& {Koen}}]{o1997}
{O'Donoghue}, D., {Lynas-Gray}, A.~E., {Kilkenny}, D., {Stobie}, R.~S., \&
  {Koen}, C. 1997, \mnras, 285, 657

\bibitem[{{O'Toole} \& {Heber}(2006)}]{otoole06}
{O'Toole}, S.~J. \& {Heber}, U. 2006, \aap, 452, 579

\bibitem[{{O'Toole} {et~al.}(2004){O'Toole}, {Heber}, \& {Benjamin}}]{otoole04}
{O'Toole}, S.~J., {Heber}, U., \& {Benjamin}, R.~A. 2004, \aap, 422, 1053

\bibitem[{{Quirion} {et~al.}(2010){Quirion}, {Podmore}, \&
  {Dupuis}}]{quirion2010}
{Quirion}, P.-O., {Podmore}, H., \& {Dupuis}, J. 2010, in American Institute of
  Physics Conference Series, Vol. 1273, American Institute of Physics
  Conference Series, ed. K.~{Werner} \& T.~{Rauch}, 554--557

\bibitem[{{Randall} {et~al.}(2005){Randall}, {Fontaine}, {Brassard}, \&
  {Bergeron}}]{rand2005}
{Randall}, S.~K., {Fontaine}, G., {Brassard}, P., \& {Bergeron}, P. 2005,
  \apjs, 161, 456

\bibitem[{{Reed} {et~al.}(2004){Reed}, {Kawaler}, {Zola}, {Jiang}, \&
  et~al.}]{reed2004}
{Reed}, M.~D., {Kawaler}, S.~D., {Zola}, S., {Jiang}, X.~J., \& et~al. 2004,
  \mnras, 348, 1164

\bibitem[{{Saffer} {et~al.}(1994){Saffer}, {Bergeron}, {Koester}, \&
  {Liebert}}]{saf94}
{Saffer}, R.~A., {Bergeron}, P., {Koester}, D., \& {Liebert}, J. 1994, \apj,
  432, 351

\bibitem[{{Seaton}(1979)}]{sea79}
{Seaton}, M.~J. 1979, \mnras, 187, 73P

\bibitem[{{Stobie} {et~al.}(1997){Stobie}, {Kawaler}, {Kilkenny}, {O'Donoghue},
  \& {Koen}}]{stobie1997}
{Stobie}, R.~S., {Kawaler}, S.~D., {Kilkenny}, D., {O'Donoghue}, D., \& {Koen},
  C. 1997, \mnras, 285, 651

\bibitem[{{Van Grootel} {et~al.}(2008){Van Grootel}, {Charpinet}, {Fontaine},
  \& {Brassard}}]{vang08}
{Van Grootel}, V., {Charpinet}, S., {Fontaine}, G., \& {Brassard}, P. 2008,
  \aap, 483, 875

\bibitem[{{Wesemael} {et~al.}(1992){Wesemael}, {Fontaine}, {Bergeron},
  {Lamontagne}, \& {Green}}]{wes92}
{Wesemael}, F., {Fontaine}, G., {Bergeron}, P., {Lamontagne}, R., \& {Green},
  R.~F. 1992, \aj, 104, 203

\end{thebibliography}

\end{document}